\documentclass[12pt]{article}
\usepackage{amsmath}
\usepackage{amssymb}
\usepackage{graphicx}
\usepackage{axodraw}
\usepackage{epsfig}
\usepackage[small]{caption}
\usepackage[dvips,usenames]{color}

\begin{document}
\baselineskip 0.6cm

\def\simgt{\mathrel{\lower2.5pt\vbox{\lineskip=0pt\baselineskip=0pt
           \hbox{$>$}\hbox{$\sim$}}}}
\def\simlt{\mathrel{\lower2.5pt\vbox{\lineskip=0pt\baselineskip=0pt
           \hbox{$<$}\hbox{$\sim$}}}}
\def\simprop{\mathrel{\lower3.0pt\vbox{\lineskip=1.0pt\baselineskip=0pt
             \hbox{$\propto$}\hbox{$\sim$}}}}

\begin{titlepage}

\begin{flushright}
UCB-PTH-11/05 \\
NSF-KITP-11-112 \\
\end{flushright}

\vskip 1.5cm

\begin{center}
{\Large \bf The Cosmological Constant in the Quantum Multiverse}

\vskip 0.7cm

{\large Grant Larsen, Yasunori Nomura, and Hannes L.L. Roberts}

\vskip 0.4cm

{\it Berkeley Center for Theoretical Physics, Department of Physics,\\
 University of California, Berkeley, CA 94720, USA}

\vskip 0.1cm

{\it Theoretical Physics Group, Lawrence Berkeley National Laboratory,
 CA 94720, USA}

\vskip 0.8cm

\abstract{Recently, a new framework for describing the multiverse has 
 been proposed which is based on the principles of quantum mechanics. 
 The framework allows for well-defined predictions, both regarding global 
 properties of the universe and outcomes of particular experiments, 
 according to a single probability formula.  This provides complete 
 unification of the eternally inflating multiverse and many worlds 
 in quantum mechanics.  In this paper we elucidate how cosmological 
 parameters can be calculated in this framework, and study the probability 
 distribution for the value of the cosmological constant.  We consider 
 both positive and negative values, and find that the observed value 
 is consistent with the calculated distribution at an order of magnitude 
 level.  In particular, in contrast to the case of earlier measure 
 proposals, our framework prefers a positive cosmological constant over 
 a negative one.  These results depend only moderately on how we model 
 galaxy formation and life evolution therein.}

\end{center}
\end{titlepage}

\section{Introduction}
\label{sec:intro}

An explanation of a small but nonzero cosmological constant is one 
of the major successes of the picture that our universe is one of 
the many different universes in which low energy physical laws take 
different forms~\cite{Weinberg:1987dv}.  Such a picture is also suggested 
theoretically by eternal inflation~\cite{Guth:1982pn} and the string 
landscape~\cite{Bousso:2000xa}.  This elegant picture, however, has 
been suffering from the predictivity crisis caused by an infinite number 
of events occurring in eternally inflating spacetime.  To make physical 
predictions, we need to deal with these infinities and define physically 
sensible probabilities~\cite{Guth:2000ka}.

Recently, a well-defined framework to describe the eternally inflating 
multiverse has been proposed based on the principles of quantum 
mechanics~\cite{Nomura:2011dt}.  In this framework, the multiverse is 
described as quantum branching processes viewed from a single ``observer'' 
(geodesic), and the probabilities are given by a simple Born-like rule 
applied to the quantum state describing the entire multiverse.  Any 
physical questions---either regarding global properties of the universe 
or outcomes of particular experiments---can be answered by using this 
single probability formula, providing {\it complete unification} of 
the eternally inflating multiverse and many worlds in quantum mechanics. 
Moreover, the state describing the multiverse is defined on the 
``observer's'' past light cones bounded by (stretched) apparent horizons; 
namely, consistent description of the {\it entire} multiverse is obtained 
in these limited spacetime regions.  This leads to a dramatic change 
of views on spacetime and gravity.

In this paper we present a calculation of the probability distribution 
of the cosmological constant in this new framework of the quantum 
multiverse.%
\footnote{For earlier studies of the cosmological constant 
 in the context of geometric cutoff measures, see 
 Refs.~[\ref{Martel:1997vi:X}~--~\ref{Bousso:2009ks:X}].}
We fix other physical parameters and ask what values of the cosmological 
constant $\Lambda$ we are likely to observe.  In Section~\ref{sec:measure} 
we begin by reviewing the proposal of Ref.~\cite{Nomura:2011dt}, and 
we then explain how cosmological parameters can be calculated in 
Section~\ref{sec:prediction}.  While the framework itself is well-defined, 
any practical calculation is necessarily approximate, since we need 
to model ``experimenters'' who actually make observations.  In our 
context, we need to consider galaxy formation and life evolution therein, 
which we will do in Section~\ref{sec:calc}.  We present the result of 
our calculation in Section~\ref{sec:cc}.  We find that, in contrast to 
the case with some earlier measures~\cite{Salem:2009eh}, the measure 
of Ref.~\cite{Nomura:2011dt} does not lead to unwanted preference for 
a negative cosmological constant---in fact, a positive value is preferred. 
We find that a simple anthropic condition based on metallicity of 
stars is sufficient to make the calculated distribution consistent 
with the observed value at an order of magnitude level.  We conclude 
in Section~\ref{sec:concl}.

Appendix~\ref{app:calc} lists formulae for galaxy formation used in 
our analysis.  Appendix~\ref{app:metal} discusses the anthropic condition 
coming from metallicity of stars.

\section{Probabilities in the Quantum Mechanical Multiverse}
\label{sec:measure}

Here we review aspects of the framework of Ref.~\cite{Nomura:2011dt} 
which are relevant to our calculation.  In this framework, the entire 
multiverse is described as a single quantum state as viewed from a single 
``observer'' (geodesic).  It allows us to make well-defined predictions 
in the multiverse (both cosmological and terrestrial), based on the 
principles of quantum mechanics.

Let us begin by considering a scattering process in usual (non-gravitational) 
quantum field theory.  Suppose we collide an electron and a positron, 
with well-defined momenta and spins: $\left| e^+ e^- \right>$ at $t = 
-\infty$.  According to the laws of quantum mechanics, the evolution 
of the state is deterministic.  In a relativistic regime, however, this 
evolution does {\it not} preserve the particle number or species, so 
we find
\begin{equation}
  \Psi(t = -\infty) = \left| e^+ e^- \right>
\quad\rightarrow\quad
  \Psi(t = +\infty) = c_e \left| e^+ e^- \right> 
    + c_\mu \left| \mu^+ \mu^- \right> + \cdots,
\label{eq:QFT-evolution}
\end{equation}
when we expand the state in terms of the free theory states (which is 
appropriate for $t \rightarrow \pm \infty$ when interactions are weak). 
The Hilbert space of the theory is (isomorphic to) the Fock space
\begin{equation}
  {\cal H} = \bigoplus_{n = 0}^{\infty} {\cal H}_{\rm 1P}^{\otimes n},
\label{eq:QFT-H}
\end{equation}
where ${\cal H}_{\rm 1P}$ is the single-particle Hilbert space.  Various 
``final states,'' $\left| e^+ e^- \right>, \left| \mu^+ \mu^- \right>, 
\cdots$, in Eq.~(\ref{eq:QFT-evolution}) arise simply because the time 
evolution operator causes ``hopping'' between different components of 
the Fock space in Eq.~(\ref{eq:QFT-H}).

The situation in the multiverse is quite analogous.  Suppose the universe 
was in an eternally inflating (quasi-de~Sitter) phase $\Sigma$ at some 
early time $t = t_0$.  In general, the evolution of this state is {\it not} 
along the axes determined by operators local in spacetime.  Therefore, 
at late times, the state is a superposition of different ``states''
\begin{equation}
  \Psi(t = t_0) = \left| \Sigma \right>
\quad\rightarrow\quad
  \Psi(t) = \sum_i c_i(t) \left| \mbox{(cosmic) configuration $i$} \right>,
\label{eq:multiverse-evolution}
\end{equation}
when expanded in terms of the states corresponding to definite semi-classical 
configurations.  The Hilbert space of the theory is (isomorphic to)
\begin{equation}
  {\cal H} = \bigoplus_{\cal M} {\cal H}_{\cal M},
\qquad
  {\cal H}_{\cal M} = {\cal H}_{{\cal M}, {\rm bulk}} 
    \otimes {\cal H}_{{\cal M}, {\rm horizon}},
\label{eq:multiverse-H}
\end{equation}
where ${\cal H}_{\cal M}$ is the Hilbert space for a fixed semi-classical 
spacetime ${\cal M}$, and consists of the bulk and horizon parts 
${\cal H}_{{\cal M}, {\rm bulk}}$ and ${\cal H}_{{\cal M}, {\rm horizon}}$. 
(The quantum states are defined on the ``observer's'' past light 
cones bounded by apparent horizons.)  The final state of 
Eq.~(\ref{eq:multiverse-evolution}) becomes a superposition 
of different semi-classical configurations because the evolution 
operator for $\Psi(t)$ allows ``hopping'' between different 
${\cal H}_{\cal M}$ in Eq.~(\ref{eq:multiverse-H}).

As discussed in detail in Ref.~\cite{Nomura:2011dt}, any physical 
question can be phrased as:\ ``Given what we know about our past light 
cone, $A$, what is the probability of that light cone to have properties 
$B$ as well?''  This probability is given by
\begin{equation}
  P(B|A) = \frac{\int\!dt \left< \Psi(t) \right| 
    {\cal O}_{A \cap B} \left| \Psi(t) \right>}
    {\int\!dt \left< \Psi(t) \right| {\cal O}_A \left| \Psi(t) \right>},
\label{eq:probability-AB}
\end{equation}
assuming that the multiverse is in a pure state $\left| \Psi(t) \right>$. 
(The mixed state case can be treated similarly.)  Here, ${\cal O}_A$ is 
the projection operator
\begin{equation}
  {\cal O}_A = \sum_i \left| \alpha_{A,i} \right> \left< \alpha_{A,i} \right|,
\label{eq:O_A}
\end{equation}
where $\left| \alpha_{A,i} \right>$ represents a set of orthonormal 
states in the Hilbert space of Eq.~(\ref{eq:multiverse-H}), i.e.\ 
possible past light cones, that satisfy condition $A$ (and similarly 
for ${\cal O}_{A \cap B}$).  Despite the fact that the $t$ integrals 
in Eq.~(\ref{eq:probability-AB}) run from $t = t_0$ to $\infty$, 
the resulting $P(B|A)$ is well-defined, since $\left| \Psi(t) 
\right>$ is continually ``diluted'' into supersymmetric Minkowski 
states~\cite{Nomura:2011dt}.

The formula in Eq.~(\ref{eq:probability-AB}) (or its mixed state 
version) can be used to answer questions both regarding global 
properties of the universe and outcomes of particular experiments. 
This, therefore, provides complete unification of the two concepts:\ 
the eternally inflating multiverse and many worlds in quantum 
mechanics~\cite{Nomura:2011dt}.%
\footnote{The claim that the multiverse and many worlds are the same 
 has also been made recently in Ref.~\cite{Bousso:2011up}, but the 
 physical picture there is very different.  Those authors argue that 
 quantum mechanics has operational meaning only under the existence 
 of causal horizons because making probabilistic predictions requires 
 decoherence with degrees of freedom outside the horizons.  Our picture 
 does not require such an extra agent to define probabilities (or 
 quantum mechanics).  The evolution of our $\Psi(t)$ is deterministic 
 and unitary.}
To predict/postdict physical parameters $x$, we need to choose $A$ to 
select the situation for ``premeasurement'' {\it without} conditioning 
on $x$.  We can then use various different values (ranges) of $x$ 
for $B$, to obtain the probability distribution $P(x)$.  In the next 
section, we discuss this procedure in more detail, in the context 
of calculating the probability distribution of the vacuum energy, 
$x = \rho_\Lambda \equiv \Lambda/8\pi G_N$.

\section{Predicting/Postdicting Cosmological Parameters}
\label{sec:prediction}

In order to use Eq.~(\ref{eq:probability-AB}) to predict/postdict 
physical parameters, we need to know the relevant properties of both 
the state $\left| \Psi(t) \right>$ (or its bulk part $\rho_{\rm bulk} 
\equiv {\rm Tr}_{\rm horizon} \left| \Psi(t) \right> \left< \Psi(t) 
\right|$) and the operators ${\cal O}_A$ and ${\cal O}_{A \cap B}$. 
Here we discuss them in turn.

In general, the state $\left| \Psi(t) \right>$ depends on the dynamics 
of the multiverse, including the scalar potential in the landscape, as 
well as the initial condition, e.g.\ at $t = t_0$.  Given limited current 
theoretical technology, this introduces uncertainties in predicting 
physical parameters.  However, there are certain cases in which 
these uncertainties are under control.  Consider $x = \rho_\Lambda$. 
We are interested only in a range a few orders of magnitude around 
$\rho_{\Lambda,{\rm obs}} \simeq (0.0024~{\rm eV})^4$~\cite{Komatsu:2010fb}, 
which is tiny compared with the theoretically expected range $-M_{\rm Pl}^4 
\simlt \rho_\Lambda \simlt M_{\rm Pl}^4$.  Therefore, unless the multiverse 
dynamics or initial condition has a special correlation with the value 
of the vacuum energy in the standard model (SM) vacua, we expect that 
the probabilities of having these vacua in $\left| \Psi(t) \right>$ 
is statistically uniform in $x$ within the range of interest.  (This 
corresponds to the standard assumption of statistical uniformity of 
the prior distribution of $\rho_\Lambda$~\cite{Weinberg:1987dv}.)  The 
distribution of $x = \rho_\Lambda$ is then determined purely by the 
dynamics {\it inside} the SM universes, i.e., the probability of 
developing experimenters who actually make observations of the 
vacuum energies.

Let us now turn to the operators ${\cal O}_A$ and ${\cal O}_{A \cap B}$. 
In order to predict the value of the vacuum energy which a given experimenter 
will observe, we need to choose ${\cal O}_A$ to select a particular 
``premeasurement'' situation for that experimenter, i.e.
\begin{equation}
  P(\rho_\Lambda)\, d\rho_\Lambda = P(B|A),
\quad
  \left\{ \begin{array}{ll}
    A: & \mbox{a particular ``premeasurement'' situation} \\
    B: & \rho_\Lambda < \mbox{vacuum energy} < \rho_\Lambda + d\rho_\Lambda,
  \end{array} \right.
\label{eq:P-rho}
\end{equation}
where $P(B|A)$ is defined in Eq.~(\ref{eq:probability-AB}).  Here, we 
have assumed that the number of SM vacua is sufficiently large for 
$\rho_\Lambda$ to be treated as continuous in the range of interest. 
In general, the specification of the premeasurement situation can be 
arbitrarily precise; for example, we can consider a particular person 
taking a particular posture in a particular room, with the tip of the 
light cone used to define $\left| \Psi(t) \right>$ located at a particular 
point in space.  In practice, however, we are interested in the vacuum 
energy ``a generic observer'' will measure.  We therefore need to relax 
the condition we impose as $A$; in other words, we need to ``coarse 
grain'' the premeasurement situation.  In fact, some coarse graining 
is always necessary when we apply the formalism to postdiction (see 
discussions in Ref.~\cite{Nomura:2011dt}).

What condition $A$ should we impose then?  To address this issue, let us 
take the semi-classical picture of the framework, discussed in Section~2 of 
Ref.~\cite{Nomura:2011dt}.  In this picture, the probability is given by
\begin{equation}
  P(B|A) = \lim_{n \rightarrow \infty} 
    \frac{{\cal N}_{A \cap B}}{{\cal N}_A},
\label{eq:semi-class-BA}
\end{equation}
where ${\cal N}_A$ is the number of past light cones that satisfy $A$ 
and are encountered by one of the $n$ geodesics emanating from randomly 
distributed points on the initial hypersurface at $t = t_0$.  (This 
is equivalent to Eq.~(\ref{eq:probability-AB}) in the regime where 
the semi-classical picture is valid.)  Since we vary only $\rho_\Lambda$, 
all the SM universes look essentially identical at early times when the 
vacuum energy is negligible.  The assumed lack of statistical correlation 
between $\rho_\Lambda$ and the multiverse dynamics then implies that 
we can consider a fixed number of geodesics emanating from a fixed 
{\it physical} volume at an early time (e.g.\ at reheating) in 
these universes, and see what fraction of these geodesics find the 
``premeasurement'' situation $A$ in each of these universes.

Given that we are focusing on the SM universes in which only the values 
of $\rho_\Lambda$ are different, it is reasonable to expect that all 
the experimenters look essentially identical for different $\rho_\Lambda$, 
at least statistically---in particular, we assume that they have similar 
sizes, masses, and lifetimes.  With this ``coarse graining,'' the 
condition $A$ can be taken, e.g., as:\ the geodesic intersects with 
the body of an experimenter at some time during their life.  In practice, 
this makes the probability proportional to the fraction of a fixed 
comoving volume at an early time that later intersects with the body 
of an observer.  Note that the details of the condition $A$ here do 
not matter for the final results---for example, we can replace the 
``body'' by ``head'' or ``nose'' without changing the results because 
its effect drops out from the normalized probability.  Thus, in this 
situation (and any situation in which condition $A$ can be formulated 
entirely in terms of things directly encountered by the geodesic), the 
semi-classical approximation to the scheme of Ref.~\cite{Nomura:2011dt} 
can be calculated as the fat geodesic measure outlined in 
Ref.~\cite{Bousso:2008hz}.%
\footnote{To our knowledge, no detailed study of the probability 
 distribution of the cosmological constant according to the fat geodesic 
 measure has been published prior to this work.}
We emphasize that the consistent quantum mechanical solution to the 
measure problem in Ref.~\cite{Nomura:2011dt} {\it forces} this choice 
on us.

We can now present the formula for $P(\rho_\Lambda)$ in a more manageable 
form.  Since the probability for one of the geodesics to intersect an 
experimenter is proportional to the number of experimenters and the 
density of geodesics, we have
\begin{equation}
  P(\rho_\Lambda) \propto 
    \sum_{a\,\in\,\mbox{\footnotesize habitable galaxies}}
    \!\!\!\!\!\!\!\!\!\!\!\!\!\!\! N_{{\rm obs},a}\,\, \rho_{{\rm geod},a},
\label{eq:P-rho_1}
\end{equation}
where $N_{{\rm obs},a}$ and $\rho_{{\rm geod},a}$ are, respectively, 
the {\it total} number of observers/experimenters and the density of 
geodesics in a ``habitable'' galaxy $a$.  Here, we have approximated 
that $\rho_{{\rm geod},a}$ is constant throughout the galaxy $a$. 
Note that since we count {\it intersections} of experimenters with 
geodesics, rather than just the number of observers (as in much previous 
work, e.g.~\cite{Martel:1997vi}), our results differ from such previous 
results by our factor of $\rho_{{\rm geod},a}$.  Our remaining task, 
then, is to come up with a scheme that can ``model'' $N_{{\rm obs},a}$ 
and $\rho_{{\rm geod},a}$ reasonably well so that the final result is 
not far from the truth.

\section{Approximating Observers}
\label{sec:calc}

In this section, we convert Eq.~(\ref{eq:P-rho_1}) into an analytic 
expression that allows us to compute $P(\rho_\Lambda)$ numerically. 
We focus on presenting the basic logic behind our arguments.  Detailed 
forms of the functions appearing below, e.g.\ $F(M,t)$ and $H(t';M,t)$, 
as well as useful fitting functions, are given in Appendix~\ref{app:calc}.

Let us begin with $N_{{\rm obs},a}$.  We assume that, at a given time $t$, 
the number of observers arising in a given galaxy $a$ is proportional to 
the total number of baryons in $a$
\begin{equation}
  \frac{d N_{{\rm obs},a}}{dt}(t) \simprop N_{B,a}(t),
\label{eq:N_obs}
\end{equation}
as long as stars are luminous.  This assumption is reasonable if the rate 
of forming observers is sufficiently small, which appears to be the case 
in our universe.  To estimate the number of baryons existing in galaxies, 
we use the standard Press-Schechter formalism~\cite{Press:1973iz}, which 
provides the fraction of matter collapsed into halos of mass larger than 
$M$ by time $t$, $F(M,t)$.  Since the amount of baryons collapsed is 
proportional to that of matter, we can use this function $F$ to estimate 
the number of observers and find%
\footnote{Note that the sign of $dF/dM$ is negative because of the definition 
 of $F$.}
\begin{equation}
  P(\rho_\Lambda) \stackrel{?}{\simprop} -\int\!\!dt\! \int\!\!dM\, 
    \frac{dF(M,t)}{dM}\, \rho_{\rm geod}(M,t).
\label{eq:P-final?}
\end{equation}

The expression of Eq.~(\ref{eq:P-final?}) does not take into account 
the fact that forming intelligent observers takes time, or that observers 
appear only when stars are luminous (which we postulate, motivated 
by the assumption that we are typical observers).  To include these 
effects, we use the extended Press-Schechter formalism~\cite{Lacey:1993iv}, 
which gives the probability $H(t';M,t)$ that a halo of mass $M$ at time 
$t$ virialized before $t'$.  The probability density $P(\rho_\Lambda)$ 
can then be written as
\begin{equation}
  P(\rho_\Lambda) \stackrel{?}{\simprop}
    - \int\!\!dt\! \int\!\!dM\, \frac{dF(M,t)}{dM}\, 
    \left\{ H(t-t_{\rm evol};M,t) - H(t-t_{\rm burn};M,t) \right\}
    \rho_{\rm geod}(M,t),
\label{eq:P-final-?-2}
\end{equation}
where $t_{\rm evol}$ and $t_{\rm burn}$ are the time needed for 
intelligent observers to evolve and the characteristic lifetime of 
stars which limits the existence of life, respectively.

The density of geodesics $\rho_{\rm geod}(M,t)$ is proportional to that 
of a dark matter halo of mass $M$ at time $t$, which is given by its 
average virial density:
\begin{equation}
  \rho_{\rm geod}(M,t) \approx \biggl( \frac{d F(M,t_*)}{dM} \biggr)^{-1}
    \int_0^{t_*}\! dt'\, \rho_{\rm vir}(t')\, \frac{d^2 F(M,t')}{dM dt'},
\label{eq:rho_geod-prim}
\end{equation}
where $t_* = {\rm min}\{ t, t_{\rm stop}(M) \}$ with $t_{\rm stop}(M)$ 
the time after which the number of halos of mass $M$ starts decreasing, 
i.e.\ when merging into larger structures dominates over formation 
of new halos: $d^2F/dMdt|_{t=t_{\rm stop}(M)} = 0$.  (For the explicit 
expression of $\rho_{\rm vir}$, see Appendix~\ref{app:calc}.)  In the 
interest of speeding up numerical calculation, we approximate this by 
the virial density at the time when the rate of matter collapsing into 
a halo of mass $M$, i.e.\ $-d^2F/dM dt$, becomes maximum:
\begin{equation}
  \rho_{\rm geod}(M,t) \approx \rho_{\rm vir}\!\left(\tau(M)\right),
\label{eq:rho_geod}
\end{equation}
where $\tau(M)$ is given by
\begin{equation}
  \left. \frac{d^3 F(M,t)}{dM dt^2} \right|_{t=\tau(M)} = 0.
\label{eq:tau-M}
\end{equation}
This approximation is indeed reasonable at the level of precision we 
are interested in: it works at the level of $20\%$ for $t \simgt 1.7 
\tau(M)$ where the contribution to $P(\rho_\Lambda)$ almost entirely 
comes from.

Finally, there will be several additional anthropic conditions 
for a halo to be able to host intelligent observers.  For example, 
the mass of a halo may have to be larger than some critical 
value $M_{\rm min}$ to efficiently form stars~\cite{Kauffmann:2002hv}, 
and smaller than $M_{\rm max}$ for the galaxy to be cooled 
efficiently~\cite{Tegmark:2005dy,Cattaneo:2006rp}.  Considering 
these factors, we finally obtain from Eqs.~(\ref{eq:P-final-?-2}) 
and (\ref{eq:rho_geod})
\begin{equation}
  P(\rho_\Lambda) = 
    - \frac{1}{N} \int_{t_{\rm evol}}^{t_{\rm f}}\!\!\!dt\! 
    \int_{M_{\rm min}}^{M_{\rm max}}\!\!\!dM\, \frac{dF(M,t)}{dM}\, 
    \left\{ H(t-t_{\rm evol};M,t) - H(t-t_{\rm burn};M,t) \right\}
    \rho_{\rm vir}\!\left(\tau(M)\right) n(M,t),
\label{eq:P-final}
\end{equation}
where $N$ is the normalization factor.  Here,
\begin{equation}
  t_{\rm f} = \left\{ \begin{array}{ll}
    \infty\, & \mbox{for } \rho_\Lambda \geq 0 \\
    t_{\rm crunch} \equiv \sqrt{\frac{\pi}{6 G_N |\rho_\Lambda|}}\, & 
      \mbox{for } \rho_\Lambda < 0,
  \end{array} \right.
\label{eq:t_f}
\end{equation}
and we have put anthropic conditions besides $M_{\rm min,max}$ in 
the form of a function $n$.  Note that $F$, $H$, and $\rho_{\rm vir}$ 
(and possibly $n$) all depend on the value of the vacuum energy 
$\rho_\Lambda$; see Appendix~\ref{app:calc}.

In summary, $(dF/dM) \left( H|_{t-t_{\rm evol}} - H|_{t- t_{\rm burn}} 
\right) n\, dM dt$ counts the (expected) number of observers in halos 
with mass between $M$ and $M+dM$ at time between $t$ and $t+dt$, and 
$\rho_{\rm vir}(\tau)$ is proportional to the density of geodesics in 
such a halo and time, and so Eq.~(\ref{eq:P-final}) gives the probability 
by counting their intersections (as in Eq.~(\ref{eq:P-rho_1})), with 
$n$ implementing some anthropic conditions.  One well-motivated origin 
for $n$ is metallicity of stars, which affects the rate of planet 
formation (see e.g.\ Refs.~\cite{Lineweaver:2000da,Fischer:2005}). 
Here we simply model this effect by multiplying some power $m$ of 
integrated star formation up to time $t - t_{\rm evol}$, which we 
assume to be proportional to the integrated galaxy formation rate 
for $M > M_{\rm min}$:
\begin{equation}
  n(M,t) \simprop 
    \bigl( F(M_{\rm min},{\rm min}\{t-t_{\rm evol},\tilde{t}_{\rm stop}\})
    - F(M,{\rm min}\{t-t_{\rm evol},\tilde{t}_{\rm stop}\}) \bigr)^m,
\label{eq:n-1}
\end{equation}
where $\tilde{t}_{\rm stop}$ is determined by $d\{ F(M_{\rm min},t') - 
F(M,t') \}/dt'|_{t'=\tilde{t}_{\rm stop}} = 0$.  (For the derivation 
of this expression, see Appendix~\ref{app:metal}.)  Motivated by 
the observation that the formation rate of certain (though not 
Earth-like) planets is proportional to the second power of host star 
metallicity~\cite{Fischer:2005}, we consider the case $m=2$, 
as well as $m=1$.%
\footnote{There is observational data for metallicity of galaxies in our 
 universe~\cite{Maiolino:2008gh}, which our crude model here does not 
 reproduce quantitatively.  However, when we straightforwardly extrapolate
 the empirical data to other universes, the same regions of the integrand 
 in Eq.~(\ref{eq:P-final}) are suppressed/enhanced so that the effect 
 on our calculation is qualitatively the same.  It must be noted, though, 
 that the strength of the effect may change; e.g.\ $P(\rho_\Lambda)$ 
 for our model with $m = 1$ is qualitatively similar to the distribution 
 obtained using the observation-motivated method with $m = 3$.}

Another possible anthropic condition comes from the fact that if 
a halo is too dense, it may not host a habitable solar system because 
of the effects of close encounters~\cite{Tegmark:1997in}.  Following 
Ref.~\cite{Tegmark:2005dy}, we assume this anthropic condition to 
take the form
\begin{equation}
  n_\star \sigma_\dagger v_\dagger \simlt \frac{1}{t_{\rm cr}},
\label{eq:anth-cond-1}
\end{equation}
where $n_\star$, $\sigma_\dagger$, $v_\dagger$, and $t_{\rm cr}$ are 
the density of stars, critical ``kill'' cross section, relative velocity 
of encounters, and some timescale relevant for the condition.  Since 
$n_\star \propto \rho_{\rm vir}$, $v_\dagger \sim v_{\rm vir} \propto 
M^{1/3} \rho_{\rm vir}^{1/6}$, and $\sigma_\dagger$ and $t_{\rm cr}$ 
are (expected to be) independent of $M$ and $\rho_{\rm vir}$, this is 
translated into
\begin{equation}
  n(M,t) = \Theta\biggl( 
    \tilde{\rho}_{\rm max} - \rho_{\rm vir}\!\left(\tau(M)\right) 
    \Bigl(\frac{M}{M_{\rm min}}\Bigr)^{2/7} \biggr),
\label{eq:n-2}
\end{equation}
where $\Theta(x)$ is the step function ($= 1$ for $x \geq 0$ and $= 0$ 
for $x < 0$), and we have normalized $M$ by $M_{\rm min}$.

The value of $\tilde{\rho}_{\rm max}$ is highly uncertain.  One way to 
estimate it is to follow Ref.~\cite{Tegmark:2005dy} and take
\begin{equation}
  n_\star \sim (1~{\rm pc})^{-3}
    \Bigl( \frac{\rho_{\rm vir}}{\rho_{\rm vir,MW}} \Bigr),
\quad
  \sigma_\dagger \sim \pi r_{\rm AU}^2,
\quad
  v_\dagger \sim v_{\rm vir} \sim 
    \sqrt{\frac{T_{\rm vir,MW}}{m_p}} 
    \Bigl( \frac{M}{M_{\rm MW}} \Bigr)^{1/3}\! 
    \Bigl( \frac{\rho_{\rm vir}}{\rho_{\rm vir,MW}} \Bigr)^{1/6},
\label{eq:n-sigma-v}
\end{equation}
where $m_p$ is the proton mass, $r_{\rm AU} \simeq 1.5 \times 
10^8~{\rm km}$ is the Sun-Earth distance, and $\rho_{\rm vir,MW} 
\sim 2 \times 10^{-26}~{\rm g/cm^3}$, $T_{\rm vir,MW} \sim 5 \times 
10^5~{\rm K}$, and $M_{\rm MW} \sim 1 \times 10^{12} M_\odot$ are 
the virial density, virial temperature, and mass of the Milky Way 
galaxy, respectively.  Using $t_{\rm cr} \sim t_{\rm evol} = 5~{\rm Gyr}$, 
Eq.~(\ref{eq:anth-cond-1}) leads to
\begin{equation}
  \tilde{\rho}_{\rm max} \sim 9 \times 10^3\, 
    \rho_{\rm vir,MW} \Bigl(\frac{M_{\rm MW}}{M_{\rm min}}\Bigr)^{2/7} 
  \sim 3 \times 10^{-22}~{\rm g/cm^3}.
\label{eq:rho_max-direct}
\end{equation}
This corresponds to the constraint from direct encounters, i.e.\ the 
orbit of a planet being disrupted by the passage of a nearby star.  There 
can also be a constraint from indirect encounters: a passing star perturbs 
an Oort cloud in the outer part of the solar system, triggering a 
lethal comet impact~\cite{Tegmark:2005dy}.  For a fixed $M$, this 
constraint can be about four orders of magnitude stronger than 
Eq.~(\ref{eq:rho_max-direct})
\begin{equation}
  \tilde{\rho}_{\rm max} \sim 3 \times 10^{-26}~{\rm g/cm^3};
\label{eq:rho_max-indirect}
\end{equation}
namely, our Milky Way galaxy may lie at the edge of allowed 
parameter space.

In our analysis below, we consider either or both of the above conditions 
Eqs.~(\ref{eq:n-1}) and (\ref{eq:n-2}).  In the real world, there are 
(almost certainly) more conditions needed for intelligent life to 
develop.  However, incorporating these conditions would likely improve 
the prediction/postdiction for $\rho_\Lambda$.  In this sense, our analysis 
may be viewed as a ``conservative'' assessment for the success of the 
framework, although it is still subject to uncertainties coming from 
the modeling of observers.

\section{Distribution of the Cosmological Constant}
\label{sec:cc}

Our modeling of observers has several parameters which need to be 
determined phenomenologically: Eq.~(\ref{eq:P-final}) contains 
$M_{\rm min}$, $M_{\rm max}$, $t_{\rm evol}$, and $t_{\rm burn}$, 
while Eq.~(\ref{eq:n-2}) contains $\tilde{\rho}_{\rm max}$.  We take 
the ``minimum'' galaxy mass appearing in Eq.~(\ref{eq:P-final}) to be
\begin{equation}
  M_{\rm min} = 2 \times 10^{11} M_\odot,
\label{eq:M_min}
\end{equation}
below which the efficiency of star formation drops 
abruptly~\cite{Kauffmann:2002hv}.  For $t_{\rm evol}$, and 
$t_{\rm burn}$, we take them approximately to be the age of 
the Earth and lifetime of the Sun, respectively:
\begin{equation}
  t_{\rm evol} = 5~{\rm Gyr},
\qquad
  t_{\rm burn} = 10~{\rm Gyr}.
\label{eq:t_obs-t_burn}
\end{equation}
In our analysis below, we use Eqs.~(\ref{eq:M_min}) and 
(\ref{eq:t_obs-t_burn}); we do not impose the constraint from galaxy 
cooling, i.e.\ we set $M_{\rm max} = \infty$.  While the values of 
these parameters are highly uncertain, our results are not very sensitive 
to these values.  The dependence of our results on them will be discussed 
at the end of this section.

\begin{figure}[t]
  \center{\includegraphics[scale=0.39]{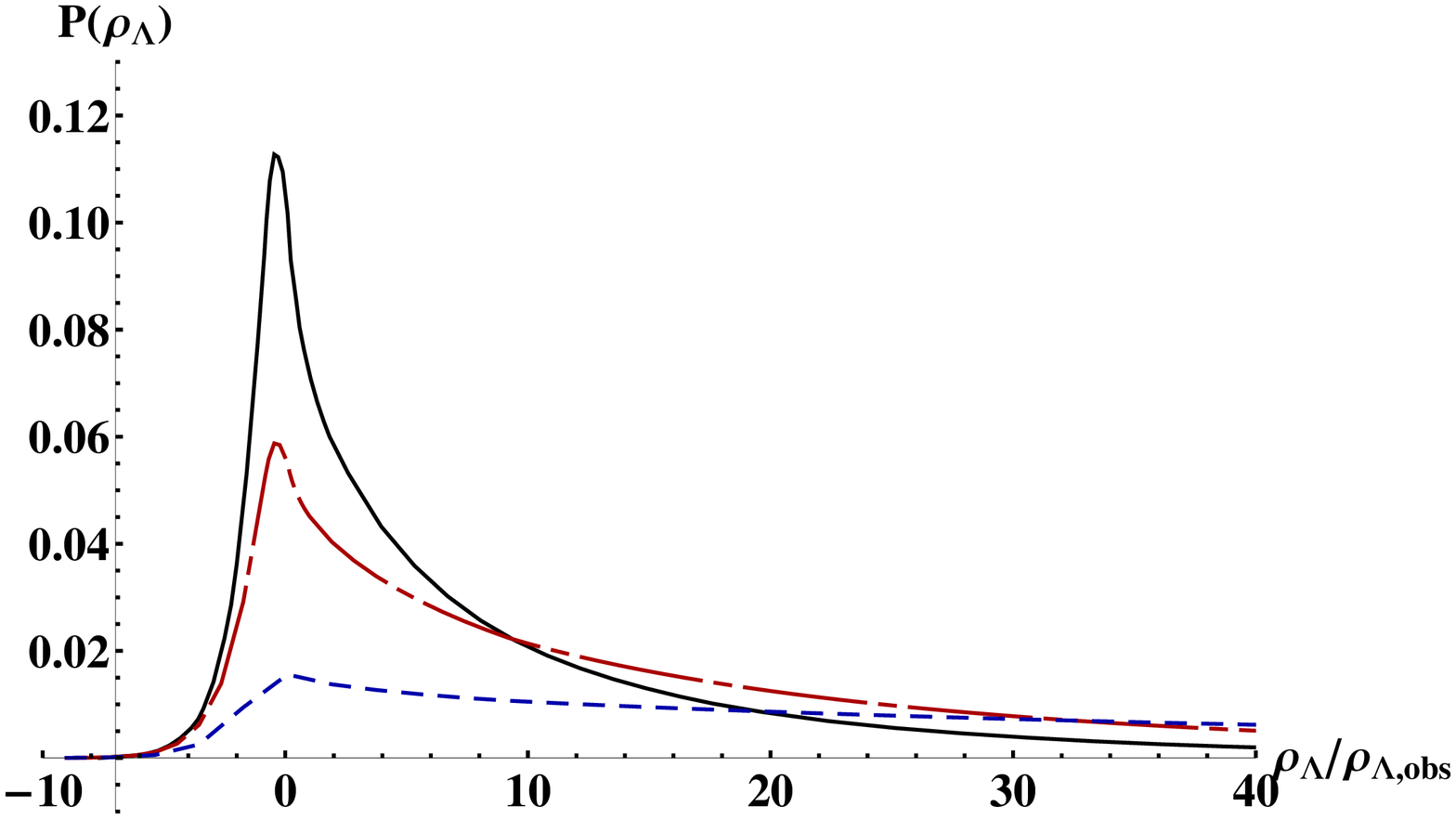}\quad
          \includegraphics[scale=0.39]{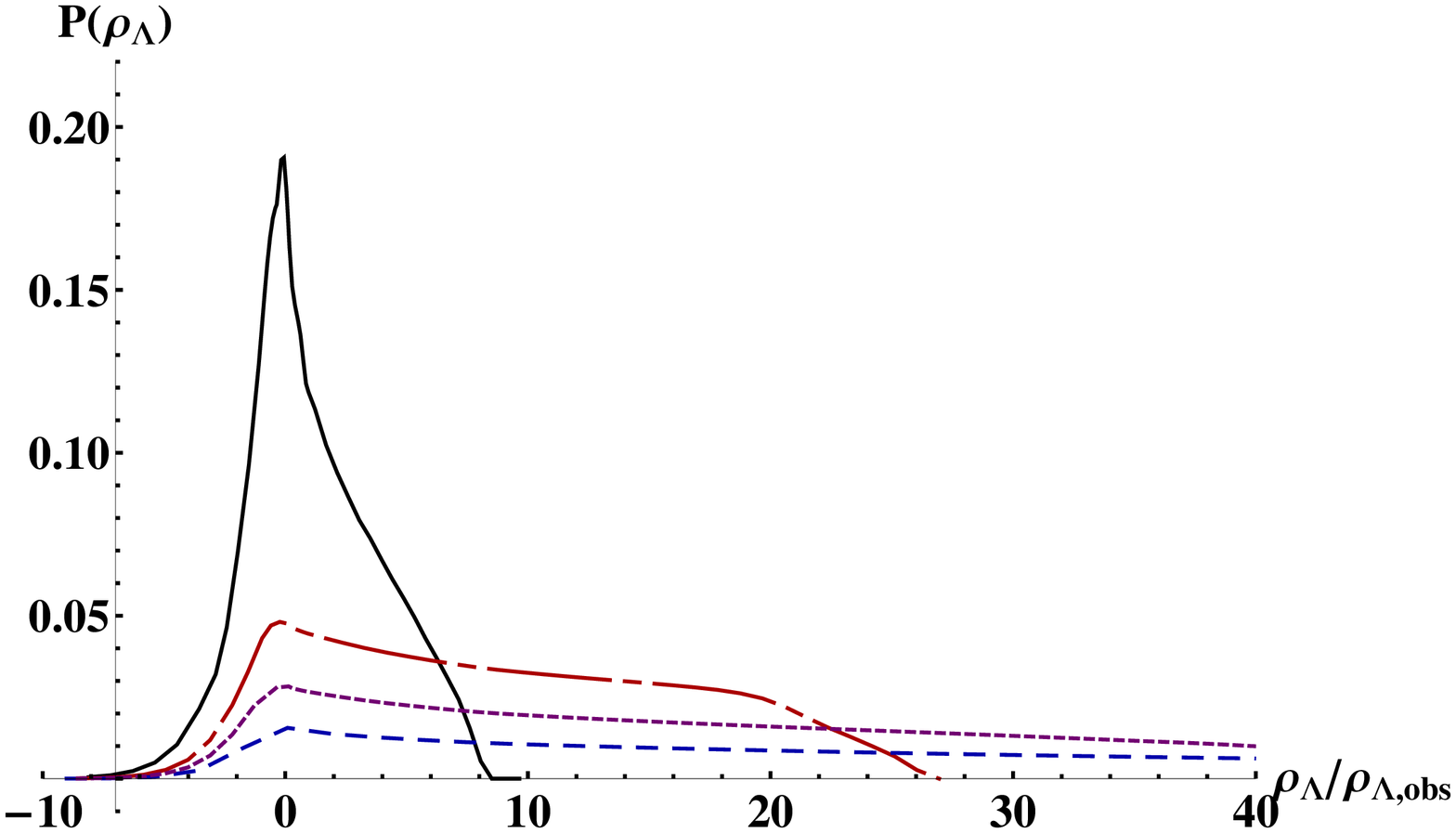}}
\caption{The normalized probability distribution of the vacuum energy 
 $P(\rho_\Lambda)$ as a function of $\rho_\Lambda/\rho_{\Lambda,{\rm obs}}$. 
 The left panel shows $P(\rho_\Lambda)$ with the metallicity condition, 
 Eq.~(\ref{eq:n-1}), with $m=0$ (i.e.\ no condition; dashed, blue), 
 $m=1$ (dot-dashed, red), and $m=2$ (solid, black).  The right panel 
 shows $P(\rho_\Lambda)$ with the upper bound $\tilde{\rho}_{\rm max}$, 
 Eq.~(\ref{eq:n-2}), with $\tilde{\rho}_{\rm max} = \infty$ 
 (i.e.\ no constraint; dashed, blue), $6 \times 10^{-26}~{\rm g/cm^3}$ 
 (dotted, purple), $4.5 \times 10^{-26}~{\rm g/cm^3}$ (dot-dashed, red), 
 and $3 \times 10^{-26}~{\rm g/cm^3}$ (solid, black).}
\label{fig:Prho-linear}
\end{figure}
In Fig.~\ref{fig:Prho-linear}, we present the normalized probability 
distribution for the vacuum energy $P(\rho_\Lambda)$ as a function of 
$\rho_\Lambda/\rho_{\Lambda,{\rm obs}}$, under several assumptions 
about the function $n$:
\begin{itemize}
\item[(i)]
``minimal'' anthropic condition: $n(M,t) = 1$
\item[(ii)]
metallicity condition: Eq.~(\ref{eq:n-1}) with $m = 1$ and $2$
\item[(iii)]
maximum virial density condition: Eq.~(\ref{eq:n-2}) with 
$\tilde{\rho}_{\rm max} = \{3 \times 10^{-26}$, $4.5 \times 10^{-26}$, 
$6 \times 10^{-26} \}~{\rm g/cm^3}$, which are $\{ 1, 1.5, 2 \}$ times 
the value in Eq.~(\ref{eq:rho_max-indirect}).
\end{itemize}
(The result with $\tilde{\rho}_{\rm max}$ given by 
Eq.~(\ref{eq:rho_max-direct}) is virtually identical to the case 
with the minimal anthropic condition.)  The left panel presents the 
effects of metallicity, showing (i) and (ii), while the right panel 
those of $\tilde{\rho}_{\rm max}$, with (i) and (iii).

Interestingly, in all cases, our predictions prefer a positive cosmological 
constant over a negative one, as opposed to the situation in earlier 
measure proposals where strong preferences to negative values have 
been found~\cite{Salem:2009eh}.  In Table~\ref{tab:rho-sign}, we provide 
the probabilities of having $\rho_\Lambda > 0$ (and $< 0$) in all six 
anthropic scenarios.  The absence of an unwanted preference towards 
negative $\rho_\Lambda$ is satisfactory, especially given that the 
measure of Ref.~\cite{Nomura:2011dt} was not devised to cure this problem. 
It comes from the fact that the present measure does not have a large 
volume effect associated with the global geometry of anti-de~Sitter space, 
which was responsible for a strong preference for negative $\rho_\Lambda$ 
in earlier, geometric cutoff measures~\cite{Salem:2009eh}.  In contrast 
with these measures, the quantum measure of Ref.~\cite{Nomura:2011dt} 
does {\it not} count the number of events; rather, it gives {\it quantum 
mechanical weights} for ``situations,'' i.e.\ quantum mechanical states 
as described from the viewpoint of a single observer (geodesic). 
The preference towards a positive value comes from the fact that 
for $\rho_\Lambda > 0$ some observers still form after vacuum energy 
domination, while for $\rho_\Lambda < 0$ it is not possible due to 
the big crunch.
\begin{table}
\begin{center}
\begin{tabular}{|c|cc|}
\hline
  & $P(\rho_\Lambda>0)$ &  $P(\rho_\Lambda<0)$ \\ \hline
  No condition & $97\%$ & $3\%$ \\
  Metallicity, $m=1$ & $87\%$ & $13\%$ \\
  Metallicity, $m=2$ & $75\%$ & $25\%$ \\
  $\tilde{\rho}_{\rm max} = 6 \times 10^{-26}~{\rm g/cm^3}$ & 
    $92\%$ & $8\%$ \\
  $\tilde{\rho}_{\rm max} = 4.5 \times 10^{-26}~{\rm g/cm^3}$ & 
    $83\%$ & $17\%$ \\
  $\tilde{\rho}_{\rm max} = 3 \times 10^{-26}~{\rm g/cm^3}$ & 
    $63\%$ & $37\%$ \\
\hline
\end{tabular}
\end{center}
\caption{The probability of observing a positive and negative cosmological 
 constant, $P(\rho_\Lambda>0)$ and $P(\rho_\Lambda<0)$, for six assumptions 
 on the anthropic condition.  In all cases, a positive value is preferred 
 over a negative one, consistent with observation.}
\label{tab:rho-sign}
\end{table}

Figure~\ref{fig:Prho-linear} shows that $P(\rho_\Lambda)$ is always 
peaked near $\rho_\Lambda = 0$, with the distribution becoming wider 
as the anthropic condition gets weaker.
\begin{figure}[t]
  \center{\includegraphics[scale=0.43]{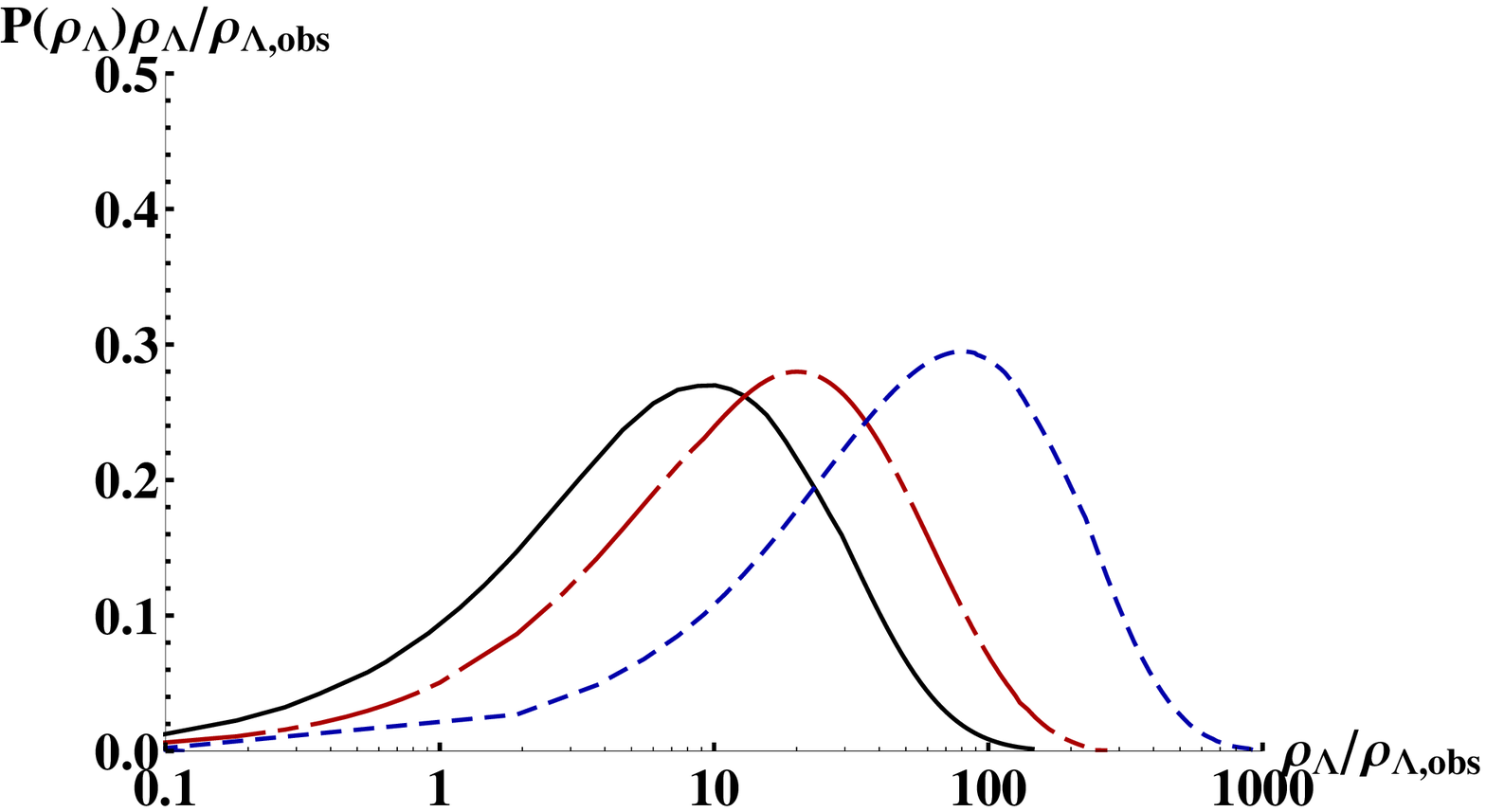}\quad
          \includegraphics[scale=0.43]{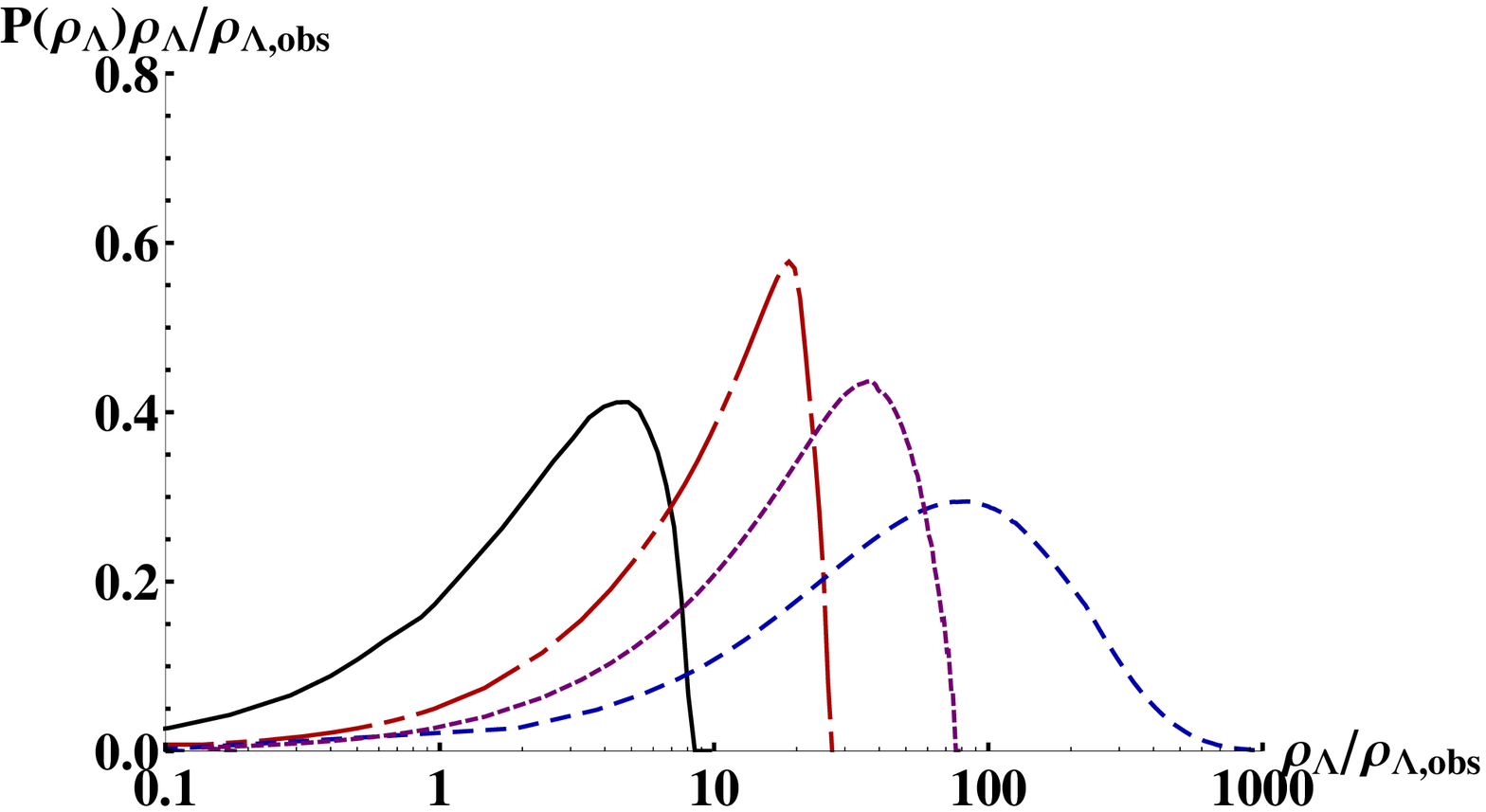}}
\caption{Same as Fig.~\ref{fig:Prho-linear}, but the horizontal axis 
 now in logarithmic scale.  To show the probability density per tenfold, 
 the vertical axis is chosen to be $\rho_\Lambda P(\rho_\Lambda) 
 / \rho_{\Lambda,{\rm obs}}$.  The distributions are normalized in 
 the region $\rho_\Lambda > 0$.}
\label{fig:Prho-log}
\end{figure}
In Fig.~\ref{fig:Prho-log}, we plot the same distributions in 
logarithmic scale for $\rho_\Lambda/\rho_{\Lambda,{\rm obs}}$, limiting 
ourselves to $\rho_\Lambda > 0$.  To show the probability density per 
tenfold, the vertical axis is chosen as $\rho_\Lambda P(\rho_\Lambda) / 
\rho_{\Lambda,{\rm obs}}$.  From these figures, we find that our 
anthropic assumptions lead to results that are consistent with the 
observed value within one or two orders of magnitude.  In particular, 
metallicity alone is enough to bring the agreement to an order of 
magnitude level.  This is because mergers, which lead to an increase 
in metallicity, are suppressed for larger values of $\rho_\Lambda$ 
due to earlier vacuum energy domination.  This result is comfortable, 
especially given that the constraint from encounters is effective only 
if $\tilde{\rho}_{\rm max}$ is close to the Milky Way value, as in 
Eq.~(\ref{eq:rho_max-indirect}).  Given our crude treatment of observers, 
we consider these results quite successful.

Finally, we discuss the sensitivity of our results to variations of 
$M_{\rm min}$, $M_{\rm max}$, $t_{\rm evol}$, and $t_{\rm burn}$, which 
can be thought of as ``systematic effects'' of our analysis. 
\begin{figure}[t]
  \center{\includegraphics[scale=0.39]{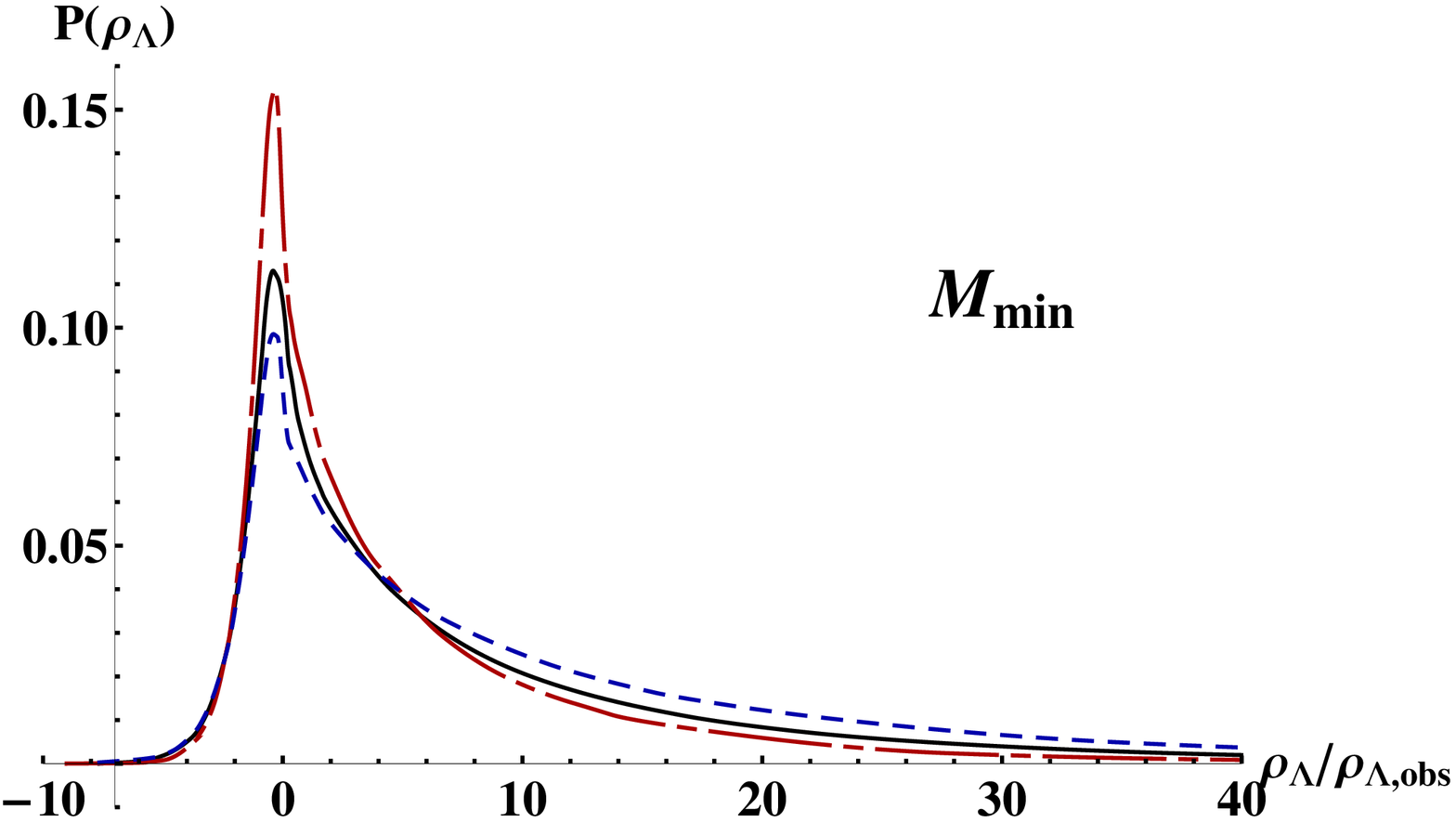}\quad
          \includegraphics[scale=0.39]{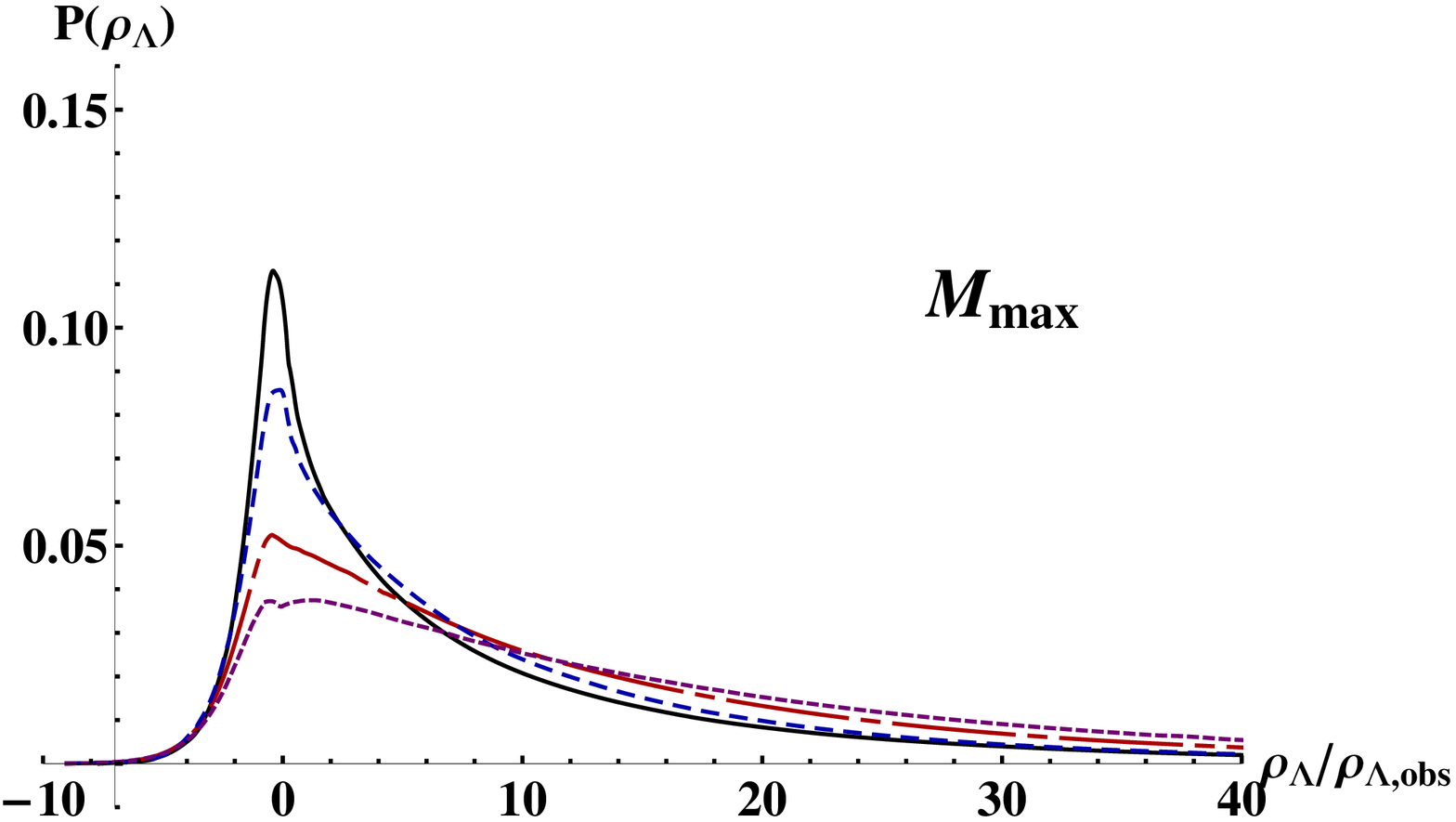}\vspace{3mm}\\
          \includegraphics[scale=0.39]{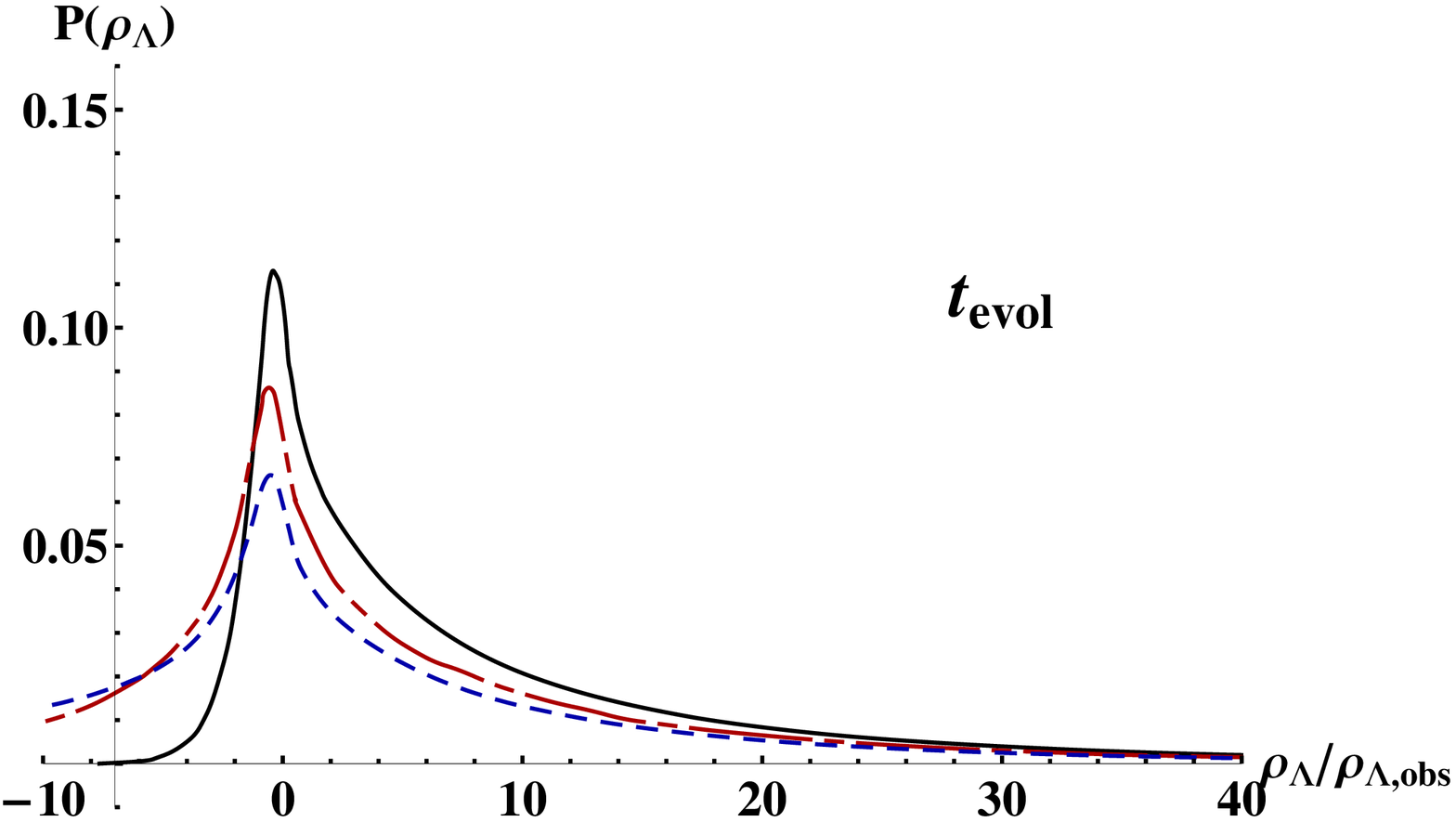}\quad
          \includegraphics[scale=0.39]{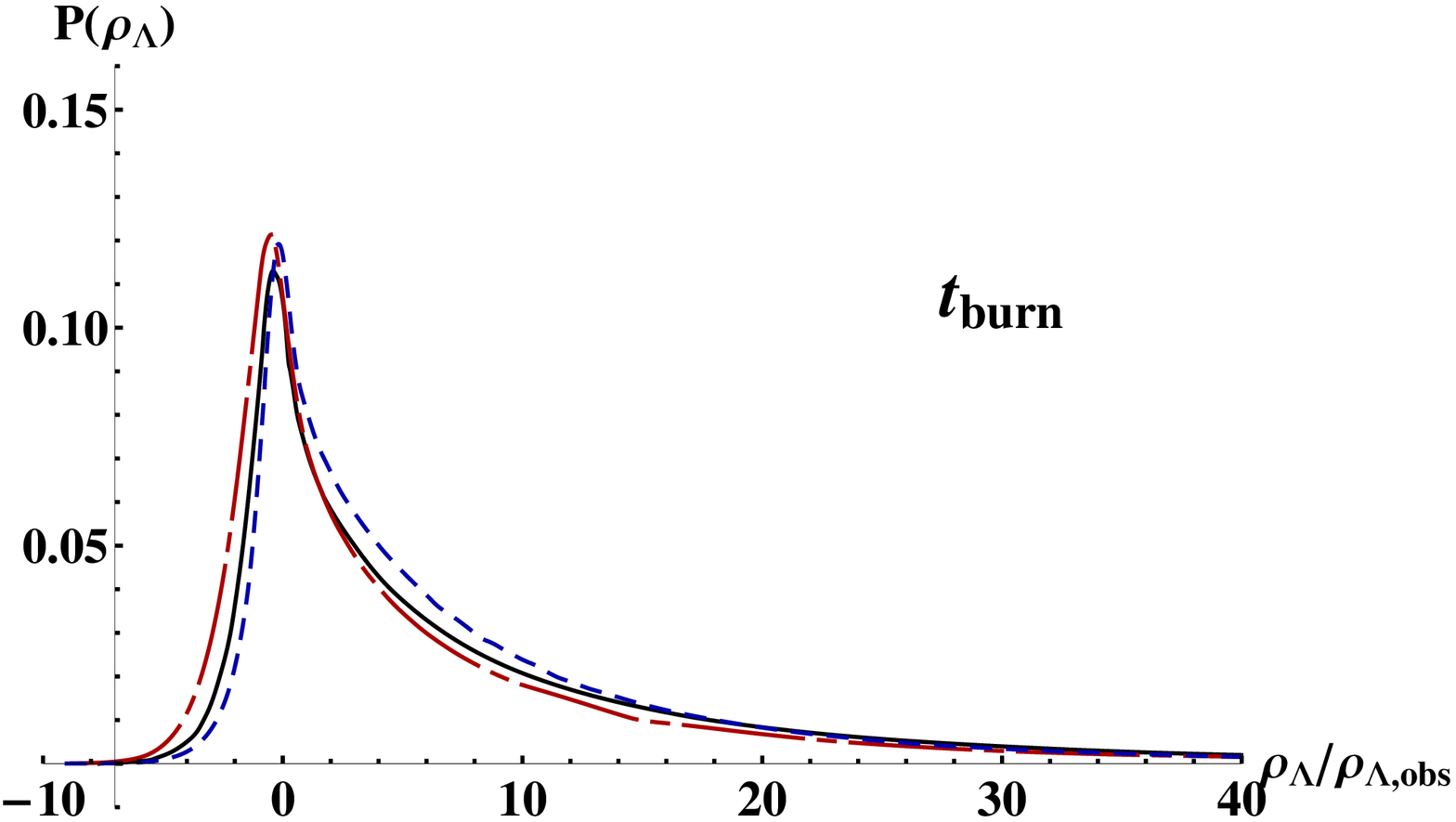}}
\caption{The normalized probability distribution $P(\rho_\Lambda)$ with 
 the metallicity condition, Eq.~(\ref{eq:n-1}), with $m=2$.  In the 
 upper-left panel, $M_{\rm min}$ is varied as $2 \times 10^{11} M_\odot$ 
 (solid, black), $6 \times 10^{11} M_\odot$ (dot-dashed, red), and 
 $0.67 \times 10^{11} M_\odot$ (dashed, blue); and in the upper-right, 
 $M_{\rm max}$ as $\infty$ (solid, black), $10^{14} M_\odot$ (dashed, 
 blue), $10^{13} M_\odot$ (dot-dashed, red), and $2 \times 10^{12} 
 M_\odot$ (dotted, purple).  The lower left and right panels vary 
 $t_{\rm evol}$ and $t_{\rm burn}$ as $\{\mbox{(solid, black), 
 (dot-dashed, red), (dashed, blue)}\} = \{5, 1, 0\}~{\rm Gyr}$ 
 and $\{10, 7, 15\}~{\rm Gyr}$, respectively.}
\label{fig:error}
\end{figure}
In Fig.~\ref{fig:error}, we show the distributions of $P(\rho_\Lambda)$ 
with the $m=2$ metallicity constraint, varying the values of $M_{\rm min}$, 
$M_{\rm max}$, $t_{\rm evol}$, and $t_{\rm burn}$, respectively.  We 
find that, while the detailed shape of $P(\rho_\Lambda)$ does change, 
our main conclusions are robust: (i) There is no strong preference 
to a negative vacuum energy; in fact, a positive value is preferred. 
(ii) The predicted distribution of $\rho_\Lambda$ is consistent with 
the observed value at an order of magnitude level with the metallicity 
constraint.

\section{Conclusions}
\label{sec:concl}

In this paper, we have studied the probability distribution of the 
cosmological constant (or the vacuum energy $\rho_\Lambda$) in the 
multiverse, using the quantum measure proposed in Ref.~\cite{Nomura:2011dt}. 
We have found that this measure does not lead to a strong preference 
for negative $\rho_\Lambda$, as opposed to earlier measures proposed 
based on geometric cutoffs, because it does not experience a large volume 
effect associated with the global geometry of anti-de~Sitter space. 
Moreover, we have found that a positive value of $\rho_\Lambda$ is 
preferred, consistent with observation.

We have found that a simple, intuitive condition based on metallicity 
is enough to reproduce the observed value of $\rho_\Lambda$ at an 
order of magnitude level.  This is comfortable because effects from 
other possible anthropic conditions, such as the ones from encounters, 
are much more sensitive to the details of the conditions.
\begin{figure}[t]
  \center{\includegraphics[scale=0.7]{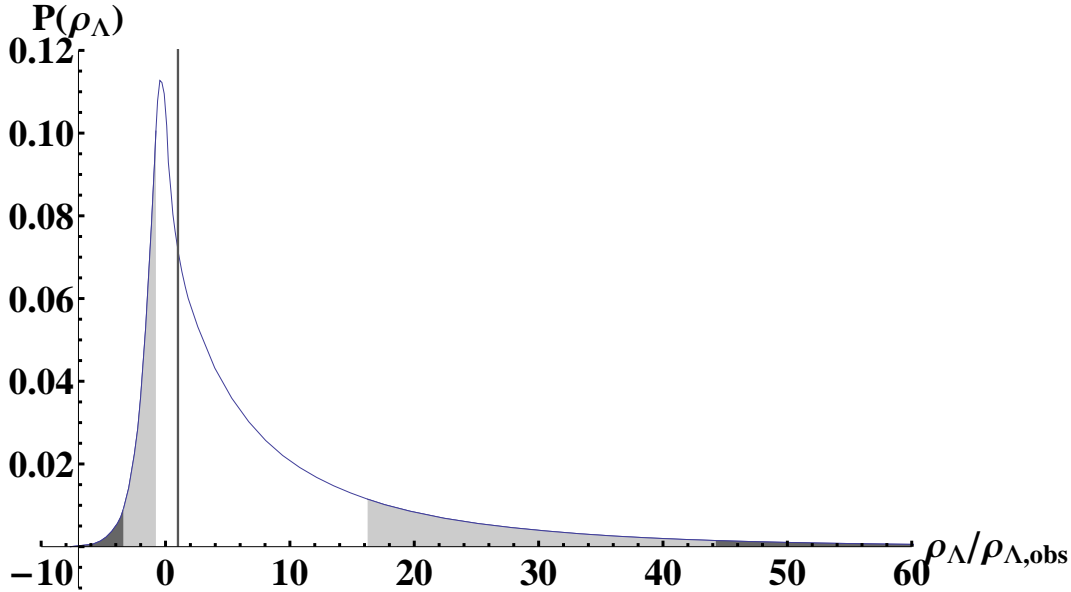}}
\caption{The normalized probability distribution $P(\rho_\Lambda)$ 
 with a metallicity condition:\ Eq.~(\ref{eq:n-1}) with $m=2$. 
 The light and dark shaded regions indicate those between $1$ and 
 $2\sigma$, and outside $2\sigma$, respectively.  The observed value 
 $\rho_\Lambda/\rho_{\Lambda,{\rm obs}} = 1$ (denoted by a vertical 
 line) is consistent with the distribution at the $1\sigma$ level.}
\label{fig:final}
\end{figure}
In Fig.~\ref{fig:final}, we present the normalized distribution 
$P(\rho_\Lambda)$ with the $m=2$ metallicity constraint, where the 
$1$ and $2\sigma$ regions are indicated.  We find that the observed 
value is consistent with the calculated distribution at the 
$1\sigma$ level.

It would be interesting to refine our analysis including more detailed 
anthropic effects, such as those of star formation.  Another possible 
extension of the analysis is to vary other cosmological parameters, such 
as the primordial density contrast $Q$ and spatial curvature $\Omega_k$ 
(at a specified time), in addition to $\rho_\Lambda$.  We plan to study 
these issues in the future.

\section*{Acknowledgments}

We thank Asimina Arvanitaki, Savas Dimopoulos, and David Pinner for 
useful discussions.  This work was supported in part by the Director, 
Office of Science, Office of High Energy and Nuclear Physics, of 
the US Department of Energy under Contract DE-AC02-05CH11231, and 
in part by the National Science Foundation under grants PHY-0855653 
and PHY05-51164.

\appendix

\section{Press-Schechter Formalism and Fitting Functions}
\label{app:calc}

The Press-Schechter function $F$ is
\begin{equation}
  F(M,t) = {\rm erfc}\left( \frac{\delta_c(t)}{\sqrt{2}\,\sigma(M,t)} \right),
\label{eq:PS-F}
\end{equation}
where $\delta_c(t)$ and $\sigma(M,t)$ are given 
by~\cite{Tegmark:2005dy,Peacock:2007cw}
\begin{equation}
  \delta_c(t) \simeq \left\{ \begin{array}{ll}
    1.629 + 0.057\, e^{-2.3 G_N \rho_\Lambda t^2}\, & 
      \mbox{for } \rho_\Lambda \geq 0 \\
    1.686 + 0.165 \left(\frac{t}{t_{\rm crunch}}\right)^{2.5} 
      + 0.149 \left(\frac{t}{t_{\rm crunch}}\right)^{11}\, & 
      \mbox{for } \rho_\Lambda < 0,
  \end{array} \right.
\label{eq:delta_c}
\end{equation}
with $t_{\rm crunch}$ defined in Eq.~(\ref{eq:t_f}), and
\begin{equation}
  \sigma(M,t) \simeq Q\, s(M)\, G(t).
\label{eq:sigma}
\end{equation}
Here, $Q$ is the primordial density contrast,
\begin{equation}
  s(M) \simeq \Bigl[ (9.1\, \mu^{-2/3})^{-0.27} 
    + \bigl\{ 50.5\, \log_{10}(834 + \mu^{-1/3}) - 92 \bigr\}^{-0.27} 
    \Bigr]^{-1/0.27},
\label{eq:s-M}
\end{equation}
where $\mu = M \xi^2 G_N^{3/2}$ with $\xi \equiv \rho_{\rm matter}/n_\gamma 
\simeq 3.7~{\rm eV}$, and
\begin{equation}
  G(t) \simeq \left\{ \begin{array}{ll}
    0.206 \frac{\xi^{4/3}}{\rho_\Lambda^{1/3}} 
        \Bigl[ \tanh^{2/3}(\frac{3}{2}H_\Lambda t) 
        \bigl\{ 1 - \tanh^{1.27}(\frac{3}{2}H_\Lambda t) \bigr\}^{0.82} & \\
    \qquad\qquad {} + 1.437 \bigl\{ 
        1 - \cosh^{-4/3}(\frac{3}{2}H_\Lambda t) \bigr\} \Bigr] & 
      \mbox{for } \rho_\Lambda \geq 0 \\
    0.549\, \xi^{4/3} G_N^{1/3} t^{2/3} 
      \Bigl[ 1 + 0.37\left(\frac{t}{t_{\rm crunch}}\right)^{2.18} \Bigr]^{-1} 
      \Bigl[ 1 - \left(\frac{t}{t_{\rm crunch}}\right)^2 \Bigr]^{-1} & 
      \mbox{for } \rho_\Lambda < 0,
  \end{array} \right.
\label{eq:G-t}
\end{equation}
where $H_\Lambda \equiv \sqrt{8\pi G_N |\rho_\Lambda|/3}$.

The function $H$ in the extended Press-Schechter formalism is given 
by~\cite{Lacey:1993iv,Bousso:2008bu}
\begin{equation}
  H(t';M,t) = -\int_{M/2}^M\! \frac{M}{M'}\, 
    \frac{d\beta}{dM'}(M',t',M,t)\, dM',
\label{eq:ePS-H}
\end{equation}
where
\begin{equation}
  \beta(M_1,t_1,M_2,t_2) = {\rm erfc}\biggl( 
    \frac{1}{Q\sqrt{2(s(M_1)^2-s(M_2)^2)}} \Bigl( 
    \frac{\delta_c(t_1)}{G(t_1)}-\frac{\delta_c(t_2)}{G(t_2)} \Bigr) \biggr),
\label{eq:beta}
\end{equation}
with $s(M)$ and $G(t)$ defined in Eqs.~(\ref{eq:s-M}) and (\ref{eq:G-t}). 

The virial density as a function of time can be fit, following 
Refs.~\cite{Tegmark:2005dy,Bousso:2008bu}, as the density evolution of 
a closed universe, according to Birkhoff's theorem.  The virial density 
is then given in terms of the density at turn-around rescaled by the 
ratio of the volumes, $\rho_{\rm vir} = (R_{\rm vir}/R_{\rm turn})^3 
\rho_{\rm turn}$.  Here, $R_{\rm vir}/R_{\rm turn} \rightarrow 2$ 
at early times ($t \ll 1/H_\Lambda$) as well as for $|\rho_\Lambda| 
\rightarrow 0$ at any fixed $t$.  For positive $\rho_\Lambda$, 
$R_{\rm vir}/R_{\rm turn} = 2/(\sqrt{3} - 1) \simeq 2.73$ at late 
times~\cite{Lahav:1991wc}, while for negative $\rho_\Lambda$, 
$R_{\rm vir}/R_{\rm turn} \rightarrow 2^{2/3}$ for $t \rightarrow 
t_{\rm crunch}$.  Our fit is given by
\begin{equation}
  \rho_{\rm vir}(t) \simeq
    \left\{ \begin{array}{ll} 
      \biggl\{ \Bigl( 18\pi^2 \rho_{\rm matter}(t) 
        \frac{\sinh^2(\frac{3}{2}H_\Lambda t)}{(\frac{3}{2}H_\Lambda t)^2} 
        \Bigr)^{1.41} 
      + (40.8\, \rho_\Lambda)^{1.41} \biggr\}^{\frac{1}{1.41}}\, & 
      \mbox{for } \rho_\Lambda \geq 0 \\
    \Bigl( 18\pi^2 \rho_{\rm matter}(t) 
      \frac{\sin^2(\frac{3}{2}H_\Lambda t)}{(\frac{3}{2}H_\Lambda t)^2} 
        \Bigr) \frac{123.6}{123.6 
      + 7 \bigl( e^{4.14\frac{t}{t_{\rm crunch}}} - 1 \bigr)}\, &
      \mbox{for } \rho_\Lambda < 0,
  \end{array} \right.
\label{eq:rho_vir}
\end{equation}
where $\rho_{\rm matter}$ is the matter energy density.  This fit is 
accurate to better than $\approx 5\%$ and $2\%$ for $\rho_\Lambda \geq 0$ 
and $< 0$, respectively.%
\footnote{For $\rho_\Lambda < 0$, the approximation leading to 
 Eq.~(\ref{eq:rho_vir}), i.e.\ $\rho_{\rm vir} \simgt \mbox{a few } 
 (\rho_{\rm matter} + \rho_\Lambda)$, breaks down for $t/t_{\rm crunch} 
 \simgt 0.8$, where we should rather use $\rho_{\rm vir} = 0$ (since 
 there is no stable structure forming).  However, since $\rho_{\rm vir}$ 
 in Eq.~(\ref{eq:rho_vir}) is small there anyway, using it up to 
 $t/t_{\rm crunch} = 1$ does not lead to a significant error.}

Finally, the time at which most of galaxies of mass $M$ forms, i.e.\ 
the solution to Eq.~(\ref{eq:tau-M}), is well approximated by the 
following fitting function:
\begin{equation}
  \tau(M)/{\rm Gyr} \simeq \frac{Q_{\rm obs}^{3/2}}{Q^{3/2}} 
  \left\{ \begin{array}{ll}
    1.880 + c_1(\alpha)\,\tilde{M} 
      + c_3(\alpha)\,\tilde{M}^3 + c_5(\alpha)\,\tilde{M}^5\, &
    \mbox{for } -10 \simlt 
      \frac{\rho_\Lambda}{\rho_{\Lambda,{\rm obs}}} < 0 \\
    c'_0(\alpha) + c'_1(\alpha)\,\tilde{M} 
      + c'_3(\alpha)\,\tilde{M}^3 + c'_5(\alpha)\,\tilde{M}^5\, & 
    \mbox{for } 0 \leq 
      \frac{\rho_\Lambda}{\rho_{\Lambda,{\rm obs}}} < 10 \\
    c''_0(\alpha) + c''_1(\alpha)\,\tilde{M} 
      + c''_3(\alpha)\,\tilde{M}^3 + c''_5(\alpha)\,\tilde{M}^5\, &
    \mbox{for } 10 \leq 
      \frac{\rho_\Lambda}{\rho_{\Lambda,{\rm obs}}} \simlt 100, \\
  \end{array} \right.
\label{eq:tau-M-fit}
\end{equation}
where $\alpha = (\rho_\Lambda/\rho_{\Lambda,{\rm obs}})(Q_{\rm obs}/Q)^3$, 
$\tilde{M} \equiv \log_{10}\frac{M}{2 \times 10^{11}M_\odot}$, and
\begin{equation}
\begin{array}{lll}
  c_1(x) &=& -0.311 + 1.276\,e^{0.827\,x} + 1.412\log_{10}\{1+|x|^{0.7}\},\\
  c_3(x) &=& 0.470 - 0.656\,e^{0.78\,x} - 0.317\log_{10}(0.2+|x|),\\
  c_5(x) &=& -0.0142 + 0.0381\,e^{0.7\,x} + 0.00822\log_{10}(0.05+|x|),\\
\\
  c'_0(x) &=& 1.880 - 0.00205\,x,\\
  c'_1(x) &=& 0.408 + 0.569\,e^{-1.01\,x} + 0.295\log_{10}(1+x),\\
  c'_3(x) &=& 0.277 - 0.251\,e^{-x} - 0.125\log_{10}(1+x),\\
  c'_5(x) &=& -0.000889 + 0.0151\,e^{-x} - 0.00220\log_{10}(1+x),\\
\\
  c''_0(x) &=& 1.880 - 0.00205\,x,\\
  c''_1(x) &=& 0.767 - 0.00293\,x -230\,x^{-4},\\
  c''_3(x) &=& -0.530 + 0.000336\,x + 0.847\,x^{-0.1},\\
  c''_5(x) &=& 0.106 - 0.0000118\,x 
             - 0.125\,x^{-0.1} - 0.0131\,\log_{10}(-5+x).
\end{array}
\label{eq:tau-M-fit_coeff}
\end{equation}
This fit is accurate to better than $\approx 5\%$ for $M \simgt 10^{11} 
M_\odot$ (but it becomes worse for smaller $M$, e.g., the accuracy 
is $\approx 12\%$ at $M \simeq 6 \times 10^{10} M_\odot$).  For 
$100 < \rho_\Lambda/\rho_{\Lambda,{\rm obs}} < 150$, we use the last 
expression of Eq.~(\ref{eq:tau-M-fit}), good to the level of $\approx 
10\%$; and for $\rho_\Lambda/\rho_{\Lambda,{\rm obs}} > 150$, we use
\begin{equation}
  \rho_{\rm vir}\!\left(\tau(M)\right)/(10^{-26}~{\rm g/cm^3}) 
  \simeq (3.66 + 0.032\,\alpha) - (1.36 + 0.0013\,\alpha)\tilde{M},
\end{equation}
which is accurate to the level of $\approx 10\%$ up to $\rho_\Lambda / 
\rho_{\Lambda,{\rm obs}} \approx 4500$.

\section{Anthropic Condition from Metallicity}
\label{app:metal}

In this appendix, we derive the function $n$ arising from the metallicity 
constraint, Eq.~(\ref{eq:n-1}).  Suppose that in a merging tree of 
a galaxy $j$ at time $t$, $j$ is found to have progenitor galaxies 
$i = 1,2,\cdots$ with varying masses $M_i$ at time $t' < t$.  Note that 
this also includes accretion, i.e.\ matter that was not part of galaxies 
of appreciable size, since $F(M=0,t) = 1$ in the Press-Schechter 
formalism, where accretion is treated as mergers of extremely tiny 
galaxies with a large galaxy.

Now, let us assume that the relative mass fraction in $j$ that 
came from $i$ and $i'$ is given by $\frac{d F(M_i,t')}{dM_i} / 
\frac{d F(M_{i'},t')}{dM_{i'}}$, i.e.\ the ratio of total amount 
of baryons at time $t'$ in galaxies of type $i$ and $i'$, respectively. 
This is true within the Press-Schechter formalism as long as $M_{i,i'} 
\ll M_j$, since then the overdensities within spherical top-hat regions 
containing masses $M_{i,i'}$ and $M_j$ are independent of each other 
at early times.  Once $M_{i,i'} \approx M_j$, the assumption is not 
justified, but in these regimes, there can only be a small amount of 
merging occurring from $i,i'$ to $j$, implying little contribution 
to metallicity.  The assumption, therefore, provides a good approximation.

Let $x_i$ ($i = 1,2,\cdots$) be the fraction of baryons in the universe 
that formed stars in halos of mass $M_i$ at time $t'$.  In our simple 
model, the star formation rate is proportional to the rate of halo 
formation for masses $M > M_{\rm min}$ and otherwise zero: $dx_i/dt' 
\propto \Theta(M_i - M_{\rm min})\, d^2 F(M_i,t')/dM_i dt'$.  The increase 
in total metal content summed over galaxies of mass $M_i$ is taken to 
be proportional to the star formation rate therein, $dx_i/dt$, so the 
increase in (linear) metallicity $dZ_i$ is
\begin{equation}
  dZ_i(t') \propto \biggl(\frac{d F(M_i,t')}{dM_i}\biggr)^{-1} 
    \dfrac{d x_i}{d t'}\, dt' 
  \propto \biggl(\frac{d F(M_i,t')}{dM_i}\biggr)^{-1} 
    \frac{d^2 F(M_i,t')}{dM_i dt'}\, \Theta(M_i - M_{\rm min})\, dt',
\label{eq:Z_i-t'}
\end{equation}
i.e. the total increase in metal content divided by the total mass.  The 
increase in metallicity of galaxy $j$ due to stars at time $t'$, then, 
has to be weighted by the relative matter fraction of galaxies $i$, as 
described above:
\begin{equation}
  d Z_j(t') = \frac{\sum_i \frac{d F(M_i,t')}{dM_i}\, dZ_i(t')} 
    {\sum_i \frac{d F(M_i,t')}{dM_i}} 
  \propto \frac{\sum_i \frac{d^2 F(M_i,t')}{dM_i dt'}\, 
    \Theta(M_i - M_{\rm min})}{\sum_i \frac{d F(M_i,t')}{dM_i}}\, dt',
\label{eq:Z_j-t'}
\end{equation}
where we must normalize to the total mass of galaxy $j$ at each time $t'$.

In the continuum limit, the sum over $i$ becomes an integral over masses. 
Therefore, the metallicity of galaxy $j$ of mass $M$ at time $t$ is
\begin{equation}
  Z(M,t) \propto \int_0^{\tilde{t}}\! dt'\, 
    \frac{\int_0^M\! dM'\, \frac{d^2F(M',t')}{dM'dt'}\, 
    \Theta(M'- M_{\rm min})}{\int_0^M\! dM'\, \frac{dF(M',t')}{dM'}} 
  = \int_0^{\tilde{t}}\! dt'\, 
    \frac{ \frac{d}{dt'} \{F(M_{\rm min},t') - F(M,t')\} }{1 - F(M,t')},
\label{eq:Z_M-t}
\end{equation}
where $\tilde t = \min\{t,\tilde{t}_{\rm stop}\}$.  The constraint that 
one cannot accumulate negative metallicity determines the timescale 
$\tilde{t}_{\rm stop}$ as a solution to
\begin{equation}
  \frac{d}{dt'} \{ F(M_{\rm min},t') - F(M,t') \} 
    \biggr|_{t' = \tilde{t}_{\rm stop}} = 0,
\label{eq:t_stop}
\end{equation}
at which time merging of galaxies of mass $M_{\rm min} < M' < M$ into 
those more massive than $M$ begins to dominate over formation of new 
galaxies in this mass region.  Since merging into larger structures 
is also occurring at earlier times, one expects that we slightly 
underestimate the metallicity.  However, in practice, the formation 
of new galaxies in this mass range and mergers into structures beyond 
are well separated in time, so a simple cutoff at $\tilde{t}_{\rm stop}$ 
is sufficient.

Now, since $F(M,t') \leq F(M,\tilde{t}_{\rm stop}) = 
{\rm erfc}(1/\sqrt{2}) \simeq 0.317$, the denominator of 
Eq.~(\ref{eq:Z_M-t}) is always between $0.68$ and $1$; in fact, it 
is very close to $1$ in most of the parameter regions.  Therefore, 
we can safely ignore the denominator of Eq.~(\ref{eq:Z_M-t}) 
and obtain
\begin{equation}
  Z(M,t) \propto \bigl( F(M_{\rm min},\tilde{t})- F(M,\tilde{t}) \bigr).
\label{eq:Z-final}
\end{equation}
In general, the probability of forming planets is 
expected to be proportional to some power $m$ of the 
metallicity~\cite{Lineweaver:2000da,Fischer:2005}.  This gives
\begin{equation}
  n(M,t) = Z(M,t-t_{\rm evol})^m,
\label{eq:app-n}
\end{equation}
which is Eq.~(\ref{eq:n-1}) in the text.


\begin{thebibliography}{99}

\bibitem{Weinberg:1987dv}
S.~Weinberg,
%``Anthropic Bound on the Cosmological Constant,''
Phys.\ Rev.\ Lett.\ {\bf 59}, 2607 (1987).
%%CITATION = PRLTA,59,2607;%%

\bibitem{Guth:1982pn}
A.~H.~Guth and E.~J.~Weinberg,
%``Could The Universe Have Recovered From A Slow First Order Phase
%Transition?,''
Nucl.\ Phys.\ B {\bf 212}, 321 (1983);
%%CITATION = NUPHA,B212,321;%%
%\bibitem{Vilenkin:1983xq}
A.~Vilenkin,
%``The Birth Of Inflationary Universes,''
Phys.\ Rev.\ D {\bf 27}, 2848 (1983);
%%CITATION = PHRVA,D27,2848;%%
%\bibitem{Linde:1986fd}
A.~D.~Linde,
%``Eternally Existing Selfreproducing Chaotic Inflationary Universe,''
Phys.\ Lett.\ B {\bf 175}, 395 (1986);
%%CITATION = PHLTA,B175,395;%%
%\bibitem{Linde:1986fc}
%A.~D.~Linde,
%``Eternal Chaotic Inflation,''
Mod.\ Phys.\ Lett.\ A {\bf 1}, 81 (1986).
%%CITATION = MPLAE,A1,81;%%

\bibitem{Bousso:2000xa}
R.~Bousso and J.~Polchinski,
%``Quantization of four-form fluxes and dynamical neutralization of the
%cosmological constant,''
JHEP {\bf 06}, 006 (2000)
[arXiv:hep-th/0004134];
%%CITATION = JHEPA,0006,006;%%
%\bibitem{Kachru:2003aw}
S.~Kachru, R.~Kallosh, A.~Linde and S.~P.~Trivedi,
%``De Sitter vacua in string theory,''
Phys.\ Rev.\ D {\bf 68}, 046005 (2003)
[arXiv:hep-th/0301240];
%%CITATION = PHRVA,D68,046005;%%
%\bibitem{Susskind:2003kw}
L.~Susskind,
%``The anthropic landscape of string theory,''
arXiv:hep-th/0302219;
%%CITATION = HEP-TH/0302219;%%
%\bibitem{Douglas:2003um}
M.~R.~Douglas,
%``The statistics of string / M theory vacua,''
JHEP {\bf 05}, 046 (2003)
[arXiv:hep-th/0303194].
%%CITATION = JHEPA,0305,046;%%

\bibitem{Guth:2000ka}
For reviews, see e.g.\
A.~H.~Guth,
%``Inflation and eternal inflation,''
Phys.\ Rept.\ {\bf 333}, 555 (2000)
[arXiv:astro-ph/0002156];
%%CITATION = PRPLC,333,555;%%
%\bibitem{Vilenkin:2006xv}
A.~Vilenkin,
%``A measure of the multiverse,''
J.\ Phys.\ A {\bf 40}, 6777 (2007)
[arXiv:hep-th/0609193];
%%CITATION = JPAGB,A40,6777;%%
%\bibitem{Winitzki:2006rn}
S.~Winitzki,
%``Predictions in eternal inflation,''
Lect.\ Notes Phys.\ {\bf 738}, 157 (2008)
[arXiv:gr-qc/0612164];
%%CITATION = LNPHA,738,157;%%
%\bibitem{Linde:2007fr}
A.~Linde,
%``Inflationary Cosmology,''
Lect.\ Notes Phys.\ {\bf 738}, 1 (2008)
[arXiv:0705.0164 [hep-th]].
%%CITATION = LNPHA,738,1;%%

\bibitem{Nomura:2011dt}
Y.~Nomura,
%``Physical Theories, Eternal Inflation, and Quantum Universe,''
arXiv:1104.2324 [hep-th].
%%CITATION = ARXIV:1104.2324;%%

\bibitem{Martel:1997vi}
\label{Martel:1997vi:X}
H.~Martel, P.~R.~Shapiro and S.~Weinberg,
%``Likely values of the cosmological constant,''
Astrophys.\ J.\  {\bf 492}, 29 (1998)
[arXiv:astro-ph/9701099];
%%CITATION = ASJOA,492,29;%%
%\bibitem{Baugh:1995hv}
G.~Efstathiou,
%``An anthropic argument for a cosmological constant,''
Mon.\ Not.\ Roy.\ Astron.\ Soc.\  {\bf 274}, L73 (1995).
%%CITATION = MNRAA,274,1049;%%

\bibitem{Garriga:1999hu}
J.~Garriga, M.~Livio and A.~Vilenkin,
%``The Cosmological constant and the time of its dominance,''
Phys.\ Rev.\  D {\bf 61}, 023503 (2000)
[arXiv:astro-ph/9906210];
%%CITATION = PHRVA,D61,023503;%%
%\bibitem{Pogosian:2006fx}
L.~Pogosian and A.~Vilenkin,
%``Anthropic predictions for vacuum energy and neutrino masses in the light of
%WMAP-3,''
JCAP {\bf 01}, 025 (2007)
[arXiv:astro-ph/0611573].
%%CITATION = JCAPA,0701,025;%%

\bibitem{Bousso:2007kq}
R.~Bousso, R.~Harnik, G.~D.~Kribs and G.~Perez,
%``Predicting the Cosmological Constant from the Causal Entropic Principle,''
Phys.\ Rev.\  D {\bf 76}, 043513 (2007)
[arXiv:hep-th/0702115];
%%CITATION = PHRVA,D76,043513;%%
%\bibitem{Cline:2007su}
J.~M.~Cline, A.~R.~Frey and G.~Holder,
%``Predictions of the causal entropic principle for environmental conditions
%of the universe,''
Phys.\ Rev.\  D {\bf 77}, 063520 (2008)
[arXiv:0709.4443 [hep-th]].
%%CITATION = PHRVA,D77,063520;%%

\bibitem{DeSimone:2008bq}
A.~De Simone, A.~H.~Guth, M.~P.~Salem and A.~Vilenkin,
%``Predicting the cosmological constant with the scale-factor cutoff
%measure,''
Phys.\ Rev.\  D {\bf 78}, 063520 (2008)
[arXiv:0805.2173 [hep-th]].
%%CITATION = PHRVA,D78,063520;%%

\bibitem{Bousso:2009ks}
\label{Bousso:2009ks:X}
R.~Bousso, L.~J.~Hall and Y.~Nomura,
%``Multiverse Understanding of Cosmological Coincidences,''
Phys.\ Rev.\  D {\bf 80}, 063510 (2009)
[arXiv:0902.2263 [hep-th]].
%%CITATION = PHRVA,D80,063510;%%

\bibitem{Salem:2009eh}
M.~P.~Salem,
%``Negative vacuum energy densities and the causal diamond measure,''
Phys.\ Rev.\  D {\bf 80}, 023502 (2009)
[arXiv:0902.4485 [hep-th]];
%%CITATION = PHRVA,D80,023502;%%
%\bibitem{Bousso:2009gx}
R.~Bousso and S.~Leichenauer,
%``Predictions from Star Formation in the Multiverse,''
Phys.\ Rev.\  D {\bf 81}, 063524 (2010)
[arXiv:0907.4917 [hep-th]];
%%CITATION = PHRVA,D81,063524;%%
%\bibitem{Bousso:2010im}
R.~Bousso, B.~Freivogel, S.~Leichenauer and V.~Rosenhaus,
%``Geometric origin of coincidences and hierarchies in the landscape,''
arXiv:1012.2869 [hep-th].
%%CITATION = ARXIV:1012.2869;%%

\bibitem{Bousso:2011up}
R.~Bousso and L.~Susskind,
%``The Multiverse Interpretation of Quantum Mechanics,''
arXiv:1105.3796 [hep-th].
%%CITATION = ARXIV:1105.3796;%%

\bibitem{Komatsu:2010fb}
E.~Komatsu {\it et al.}  [WMAP Collaboration],
%``Seven-Year Wilkinson Microwave Anisotropy Probe (WMAP) Observations:
%Cosmological Interpretation,''
Astrophys.\ J.\ Suppl.\  {\bf 192}, 18 (2011)
[arXiv:1001.4538 [astro-ph.CO]].
%%CITATION = APJSA,192,18;%%

\bibitem{Bousso:2008hz}
R.~Bousso, B.~Freivogel and I.~S.~Yang,
%``Properties of the scale factor measure,''
Phys.\ Rev.\  D {\bf 79}, 063513 (2009)
[arXiv:0808.3770 [hep-th]].
%%CITATION = PHRVA,D79,063513;%%

\bibitem{Press:1973iz}
W.~H.~Press and P.~Schechter,
%``Formation of galaxies and clusters of galaxies by selfsimilar gravitational
%condensation,''
Astrophys.\ J.\  {\bf 187}, 425 (1974).
%%CITATION = ASJOA,187,425;%%

\bibitem{Lacey:1993iv}
C.~Lacey and S.~Cole,
%``Merger rates in hierarchical models of galaxy formation,''
Mon.\ Not.\ Roy.\ Astron.\ Soc.\  {\bf 262}, 627 (1993).
%%CITATION = MNRAA,262,627;%%

\bibitem{Kauffmann:2002hv}
G.~Kauffmann {\it et al.}  [SDSS Collaboration],
%``The Dependence of star formation history and internal structure on stellar
%mass for 10**5 low-redshift galaxies,''
Mon.\ Not.\ Roy.\ Astron.\ Soc.\  {\bf 341}, 54 (2003)
[arXiv:astro-ph/0205070].
%%CITATION = MNRAA,341,54;%%

\bibitem{Tegmark:2005dy}
M.~Tegmark, A.~Aguirre, M.~J.~Rees and F.~Wilczek,
%``Dimensionless constants, cosmology and other dark matters,''
Phys.\ Rev.\  D {\bf 73}, 023505 (2006)
[arXiv:astro-ph/0511774].
%%CITATION = PHRVA,D73,023505;%%

\bibitem{Cattaneo:2006rp}
A.~Cattaneo, A.~Dekel, J.~Devriendt, B.~Guiderdoni and J.~Blaizot,
%``Modelling the galaxy bimodality: shutdown above a critical halo mass,''
Mon.\ Not.\ Roy.\ Astron.\ Soc.\  {\bf 370}, 1651 (2006)
[arXiv:astro-ph/0601295].
%%CITATION = MNRAA,370,1651;%%

\bibitem{Lineweaver:2000da}
C.~H.~Lineweaver,
%``An estimate of the age distribution of terrestrial planets in the universe:
%quantifying metallicity as a selection effect,''
Icarus {\bf 151}, 307 (2001)
[arXiv:astro-ph/0012399].
%%CITATION = ASTRO-PH/0012399;%%

\bibitem{Fischer:2005}
D.~A.~Fischer and J.~Valenti,
%``The planet-metallicity correlation,''
Astrophys.\ J.\ {\bf 622}, 1102 (2005);
%\bibitem{Grether:2007}
D.~Grether and C.~H.~Lineweaver,
%``The metallicity of stars with close companions,''
Astrophys.\ J.\ {\bf 669}, 1220 (2007).

\bibitem{Maiolino:2008gh}
R.~Maiolino {\it et al.},
%``AMAZE. I. The evolution of the mass-metallicity relation at z>3,''
Astron.\ Astrophys.\  {\bf 488}, 463 (2008)
[arXiv:0806.2410 [astro-ph]];
%%CITATION = AAEJA,488,463;%%
%\bibitem{Behroozi:2010rx} 
P.~S.~Behroozi, C.~Conroy and R.~H.~Wechsler,
%``A Comprehensive Analysis of Uncertainties Affecting the Stellar Mass-Halo Mass Relation for 0<z<4,''
Astrophys.\ J.\  {\bf 717}, 379 (2010)
[arXiv:1001.0015 [astro-ph.CO]].
%%CITATION = ARXIV:1001.0015;%%

\bibitem{Tegmark:1997in}
M.~Tegmark and M.~J.~Rees,
%``Why is the CMB fluctuation level 10**(-5)?,''
Astrophys.\ J.\  {\bf 499}, 526 (1998)
[arXiv:astro-ph/9709058].
%%CITATION = ASJOA,499,526;%%

\bibitem{Peacock:2007cw}
J.~A.~Peacock,
%``Testing anthropic predictions for Lambda and the CMB temperature,''
Mon.\ Not.\ Roy.\ Astron.\ Soc.\  {\bf 379}, 1067 (2007)
[arXiv:0705.0898 [astro-ph]].
%%CITATION = MNRAA,379,1067;%%

\bibitem{Bousso:2008bu}
R.~Bousso and S.~Leichenauer,
%``Star Formation in the Multiverse,''
Phys.\ Rev.\  D {\bf 79}, 063506 (2009)
[arXiv:0810.3044 [astro-ph]].
%%CITATION = PHRVA,D79,063506;%%

\bibitem{Lahav:1991wc}
O.~Lahav, P.~B.~Lilje, J.~R.~Primack and M.~J.~Rees,
%``Dynamical effects of the cosmological constant,''
Mon.\ Not.\ Roy.\ Astron.\ Soc.\  {\bf 251}, 128 (1991).
%%CITATION = MNRAA,251,128;%%

\end{thebibliography}
\end{document}